\documentclass[12pt]{article}
\usepackage{newtxtext,newtxmath}

\usepackage{graphicx}
\usepackage{caption}
\usepackage[linesnumbered,ruled,vlined]{algorithm2e}  % 新添加的伪代码包
\usepackage[letterpaper,margin=1in]{geometry}

\renewenvironment{abstract}
	{\quotation}
	{\endquotation}

\date{}

\makeatletter
\renewcommand{\fnum@figure}{\textbf{Figure \thefigure}}
\renewcommand{\fnum@table}{\textbf{Table \thetable}}
\makeatother

\usepackage{scicite}

\usepackage{url}

\def\scititle{
	Meta-training of diffractive meta-neural networks for super-resolution direction of arrival estimation
}
\title{\bfseries \boldmath \scititle}

\author{
	Songtao Yang$^{1,\dagger}$,
	Sheng Gao$^{1,\dagger}$,
	Chu Wu$^{1}$,
	Zejia Zhao$^{1}$,
	Haiou Zhang$^{1}$,
	Xing Lin$^{1,2,\ast}$\and
	\small$^{1}$Department of Electronic Engineering, Tsinghua University, Beijing, 100084, China.\and
	\small$^{2}$Beijing National Research Center for Information Science and Technology, Tsinghua University,\\
	\small Beijing, 100084, China.\and
	\small$^\ast$Corresponding author. Email: lin-x@tsinghua.edu.cn\and
	\small$^\dagger$These authors contributed equally to this work.
}

\begin{document} 

% Insert the title and author list
\maketitle

% Abstract, in bold
% There are strict length limits, and not all formats have abstracts.
% Consult the journal instructions to authors for details.
% Do not cite any references in the abstract.
\begin{abstract} \bfseries \boldmath

Diffractive neural networks leverage the high-dimensional characteristics of electromagnetic (EM) fields for high-throughput computing. However, the existing architectures face challenges in integrating large-scale multidimensional metasurfaces with precise network training and haven't utilized multidimensional EM field coding scheme for super-resolution sensing. Here, we propose diffractive meta-neural networks (DMNNs) for accurate EM field modulation through metasurfaces, which enable multidimensional multiplexing and coding for multi-task learning and high-throughput super-resolution direction of arrival estimation. DMNN integrates pre-trained mini-metanets to characterize the amplitude and phase responses of meta-atoms across different polarizations and frequencies, with structure parameters inversely designed using the gradient-based meta-training. For wide-field super-resolution angle estimation, the system simultaneously resolves azimuthal and elevational angles through x and y-polarization channels, while the interleaving of frequency-multiplexed angular intervals generates spectral-encoded optical super-oscillations to achieve full-angle high-resolution estimation. Post-processing lightweight electronic neural networks further enhance the performance. Experimental results validate that a three-layer DMNN operating at 27 GHz, 29 GHz, and 31 GHz achieves $\sim7\times$ Rayleigh diffraction-limited angular resolution (0.5$^\circ$), a mean absolute error of 0.048$^\circ$ for two incoherent targets within a ±11.5$^\circ$ field of view, and an angular estimation throughput an order of magnitude higher (1917) than that of existing methods. The proposed architecture advances high-dimensional photonic computing systems by utilizing inherent high-parallelism and all-optical coding methods for ultra-high-resolution, high-throughput applications.

\end{abstract}

\subsection*{Introduction}
% The first paragraph of any Science paper does NOT have a heading
% Nor is it indented
\noindent
Photonic computing has re-emerged as a promising alternative computing paradigm, offering ultrafast computation, high parallel computing capabilities, and low operational power consumption \cite{mcmahon2023physics,he2022higherdimensional}. These unique features enable photonic computing systems to facilitate the applications in diverse fields such as deep learning \cite{xu202111,feldmann2021parallel}, edge perception \cite{sludds2022delocalized}, and signal processing \cite{dong2023higherdimensional,zhang2024systemonchip,xu2024analog}. Recent research has extensively explored two complementary principal avenues of photonic computing, including on-chip integrated systems \cite{bogaerts2020programmable,chen2023allanalog,xu2024largescale,hua2025integrated,ahmed2025universal} and free-space diffractive architectures \cite{lin2018alloptical,hu2024diffractive,li2024nonlinear,yildirim2024nonlinear,yan2024eeg,zheng2023dual,qu2022alldielectric}. Compared with integrated photonic computing systems, the diffractive photonic computing systems, especially with diffractive deep neural network (D\textsuperscript{2}NN) architectures \cite{lin2018alloptical} have exceptional features in large-scale 3D interconnectivity and parallel processing capability. Such systems have demonstrated remarkable versatility across multiple electromagnetic (EM) spectra, including visible \cite{chen2021diffractive,qu2022alldielectric,duan2023optical,chen2024superresolution}, infrared \cite{zhou2021largescale}, and microwave bands \cite{qian2020performing,qian2022dynamic}.

The information processing capability of D\textsuperscript{2}NN can be substantially enhanced by utilizing metasurfaces as the diffractive modulation layers \cite{wang2024matrix,huang2023diffraction,liu2022programmable}. The feature size of metasurface structure is much smaller than the working wavelength, which allows EM waves to be manipulated across multiple dimensions based on various optical effects \cite{yu2011light}. When integrating into D\textsuperscript{2}NNs, conventional modeling methods typically ignore amplitude modulation, and only the phase modulation of each meta-atom is characterized during the network training. Therefore, a look-up table can be used to find the most suitable design parameters from pre-characterized meta-atoms. However, this strategy limits the full exploitation of the high-dimensional modulation (HDM) property of metasurfaces.

The limitation of modulation dimension can be addressed by using the deep learning-based inverse design methods for metasurfaces~\cite{ma2021deep,xiong2024deep,qian2025guidance,qian2025progress,jiang2021neural,chi2025neural,ma2022pushing,su2025multidimensional,naseri2021generative,jia2024dynamic,wu2025general}, which offer the capability of learning complex mappings between structure and response. These approaches significantly expand the design space and enable direct mapping between structural parameters and multi-dimensional EM behavior. Despite their promise, integrating such metasurfaces into D\textsuperscript{2}NNs remains challenging in terms of network training accuracy and efficiency. The separation of the task training for optical modulation parameters and metasurface inverse design for structure parameters places fundamental limitations on the training accuracy and HDM feasibility. The required training datasets and the network size for deep learning, which scale exponentially with the number of design parameters, substantially limit the training efficiency.

Recent studies further demonstrate that the D\textsuperscript{2}NN, through the design of a super-oscillatory system response~\cite{chen2019superoscillation}, enables all-optical super-resolution sensing with resolution exceeding the diffraction limit~\cite{gao2024superresolution,chen2024superresolution}. This advancement facilitates super-resolution imaging applications~\cite{chen2024superresolution} as well as low-latency integrated sensing and communication systems~\cite{gao2024superresolution}. Among these techniques, the direction-of-arrival (DOA) estimation, which determines the angular orientation of EM field sources, constitutes a critical requirement for enabling diverse applications in communication systems, radar technology, and positioning networks~\cite{huang2024comprehensive,weiwang2022highprecision,chen2025integrated,gao2024superresolution}. However, the all-optical sensing accuracy and confidence value of existing systems reduce with the increase of sensing resolution and task complexity for multiple targets. Especially, the estimation confidence value substantially reduces at the angular interval boundary for the application of DOA. All-optical multi-dimensional encoding scheme hasn't been utilized to enhance the sensing accuracy and resolution across the entire field of view (FOV) and angular region of interest.

To address these issues, we propose to construct diffractive meta-neural networks (DMNNs) by integrating internal resonated meta-atoms for HDM at each diffractive modulation layer and apply them for super-resolution DOA estimation with multi-frequency coding. The meta-training process is developed to jointly optimize meta-atom structure parameters and system optical modulation parameters and enable high-accuracy modeling and training of DMNNs to accomplish the desired tasks. This is achieved by constructing the pre-trained mini-metanet with Fourier feature multilayer perceptrons (FF-MLPs)~\cite{tancik2020fourier} architecture to learn the accurate mapping between meta-atom structure parameters and their amplitude and phase responses across different polarizations and frequencies, which are then utilized during the gradient-based training for DOA estimation. By leveraging orthogonal polarization multiplexing, the network can simultaneously estimate azimuthal and elevational angles. In addition, through frequency multiplexing, strengthened with the optional post-processing lightweight electronic neural network as the estimator, it enables accurate angular estimation over a wide FOV with further enhanced angular resolutions beyond the diffraction limits. Experimental results demonstrate that a three-layer DMNN operating at 27~GHz, 29~GHz, and 31~GHz achieves a mean absolute error of 0.028$^\circ$ for a single source and 0.048$^\circ$ for two incoherent sources, with angular estimation throughput ($AET$) an order of magnitude higher than state-of-the-art angular estimation method~\cite{gao2024superresolution}.

\subsection*{Results}
\subsubsection*{Design of DMNN architecture for DOA estimation}

\begin{figure}
  \includegraphics[width=\linewidth]{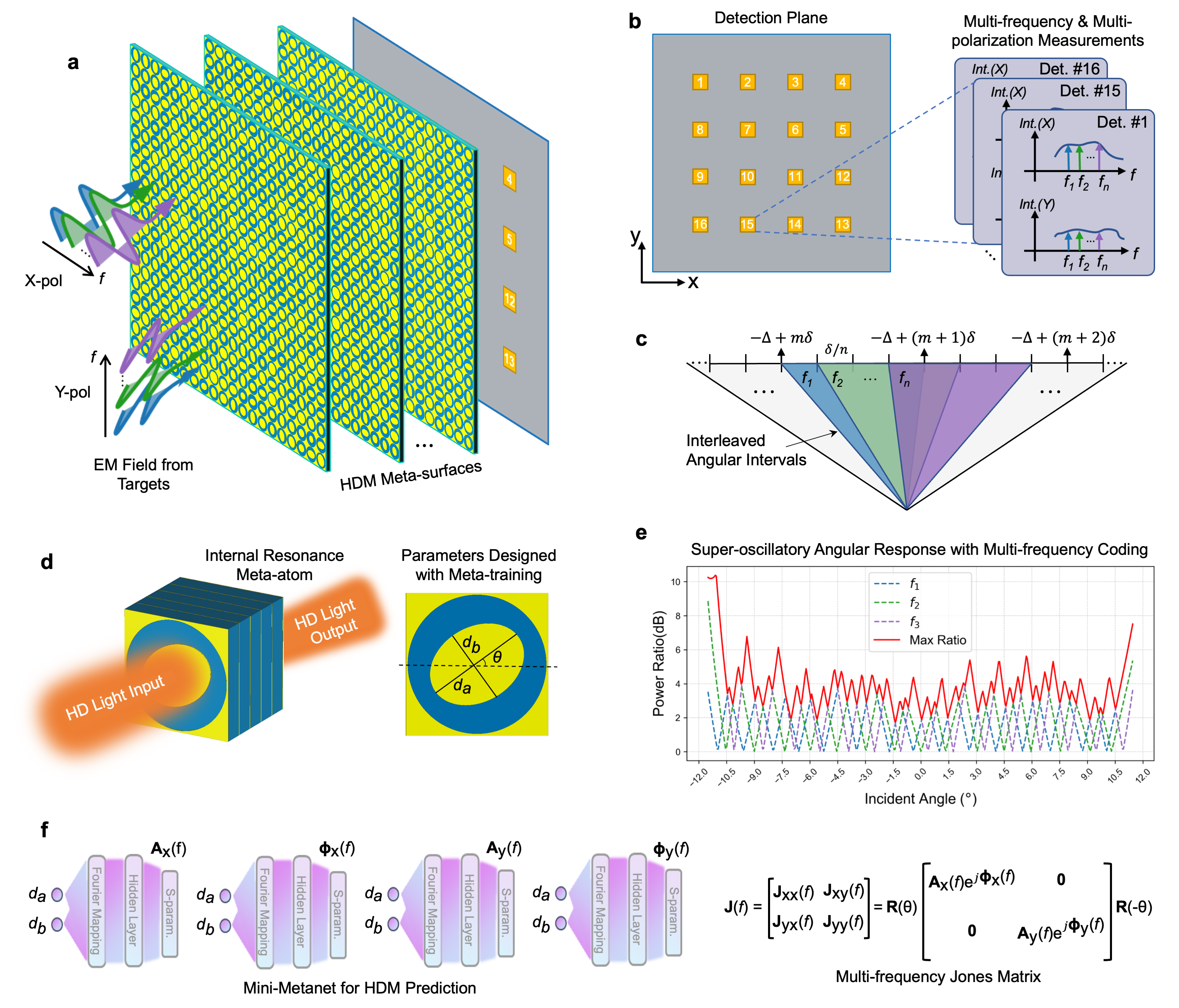}
  \caption{\textbf{DMNN for super-resolution DOA estimation.} \textbf{(a, b)} DMNN directly processes the multi-dimensional EM field from targets by measuring the intensity in x- and y-polarizations and across multiple frequency channels at the detection plane. \textbf{c} An all-optical multi-frequency coding strategy is implemented at the $m$-th angular interval by interleaving the angular intervals corresponding to different frequencies. \textbf{d} An internal resonance meta-atom for HDM is employed to process high-dimensional (HD) input light (left). It comprises stacked ellipse-shaped metals within a surrounding metallic frame with learnable parameters of $d_a, d_b$, and $\theta$ (right). \textbf{e} System super-oscillatory angular response at each frequency (dashed lines) and after combination with all-optical multi-frequency coding (solid lines). \textbf{f} Pre-train mini-metanets with FF-MLP architecture for characterizing HDM of meta-atoms.}
\label{fig1}
\end{figure}

Fig.~\ref{fig1}\textbf{a} illustrates the overall architecture of the proposed DMNN designed for super-resolution DOA estimation by directly processing multi-dimensional EM fields with a frequency coding scheme (see Methods and Supplementary Note 1 $\&$ 2). The detection plane contains 16 detection regions, as labeled in Fig.~\ref{fig1}\textbf{b}. An “S”-shaped detector arrangement, rather than a simple left-to-right sequence, is employed to minimize local field discontinuities between adjacent angular intervals. Furthermore, a delicate angular offset is introduced to interleave frequency-dependent estimation at each angular interval, as illustrated in Fig.~\ref{fig1}\textbf{c}. For the FOV setting of $[-\Delta, \Delta]$, the $m$-th angular interval $[-\Delta + m\delta, -\Delta + (m+1)\delta]$, with $\delta$ representing the angular interval width, can be further divided into $n$ sub-angular intervals by multiplexing $n$ working frequencies. Thus, the $n_1$-th sub-angular interval of the $m$-th angular interval, corresponding to the $n_1$-th working frequency, with $n_1 = 1,...,n$, can be represented as $[-\Delta + m\delta + \delta*(n_1-1)/n, -\Delta + m\delta + \delta*n_1/n]$. This design facilitates the fusion of multi-frequency detection results, thereby improving the angular estimation resolution. When a target emits EM waves, the system detects the energy distributions under x-polarization at the designated frequencies. These are synthesized to enhance azimuthal angle estimation (Fig.~\ref{fig1}\textbf{b}). Similarly, the y-polarization responses are utilized for elevational angle estimation.

The DMNN integrates a large number of internal resonance meta-atoms at each modulation layer (see Fig.~\ref{fig1}\textbf{d} and Methods), which comprise six stacked ellipse-shaped metal layers, each within a surrounding metallic frame, with learnable design parameters of $d_a, d_b$, and $\theta$. The dielectric material layers are utilized as the support layer between the metal layers. The EM responses of meta-atoms are represented by the multi-frequency Jones matrix, a complex-valued matrix comprising four elements, i.e., $\mathbf{J}_{xx}(\emph{f}), \mathbf{J}_{xy}(\emph{f}), \mathbf{J}_{yx}(\emph{f}), \mathbf{J}_{yy}(\emph{f})$ (see Fig.~\ref{fig1}\textbf{f}), each of which is a vector with a size of $n$ representing different frequencies. The matrix elements are calculated by characterizing the amplitude and phase modulation of $S_{21}$ parameters of the meta-atom at x- and y-polarizations for different frequencies, i.e., $\mathbf{A}_{x}(\emph{f})$, $\mathbf{\Phi}_{x}(\emph{f})$, $\mathbf{A}_{y}(\emph{f})$, $\mathbf{\Phi}_{y}(\emph{f})$, as well as the rotation matrices $\mathbf{R}(\theta)$ and $\mathbf{R}(-\theta)$ (see Fig.~\ref{fig1}\textbf{f}, right). The $S_{21}$ parameters of the meta-atom, represented in the form of four vectors, are predicted by training four FF-MLP-based mini-metanets (see Fig.~\ref{fig1}\textbf{f}, left). The high-dimensional EM field is decomposed into independent channels of different dimensions, which are propagated in parallel through the system to accomplish the task and map the computational results to detection regions at the output plane. The detected intensities of the output EM field at different frequencies and x- and y-polarizations are acquired for estimating the azimuthal and elevational angles of targets. Fig.~\ref{fig1}\textbf{e} presents the system's super-oscillatory angular responses, measured by calculating the power ratio of two detectors with the largest and second-largest powers among the 16 detectors, at three different frequencies of $f_1, f_2, f_3$, and after the frequency coding of these three different frequencies. The dashed lines in different colors represent the system angular responses at individual frequencies, while the solid red line indicates the result of combining multiple frequencies based on the maximum power ratio. Compared to single-frequency detection, the proposed multi-frequency all-optical coding strategy enables faster system angular response and achieves higher resolution and more accurate DOA estimation within the same FOV.

\subsubsection*{Characterize meta-atom using FF-MLP-based mini-metanets}

\begin{figure}
  \centering
  \includegraphics[width=0.9\linewidth]{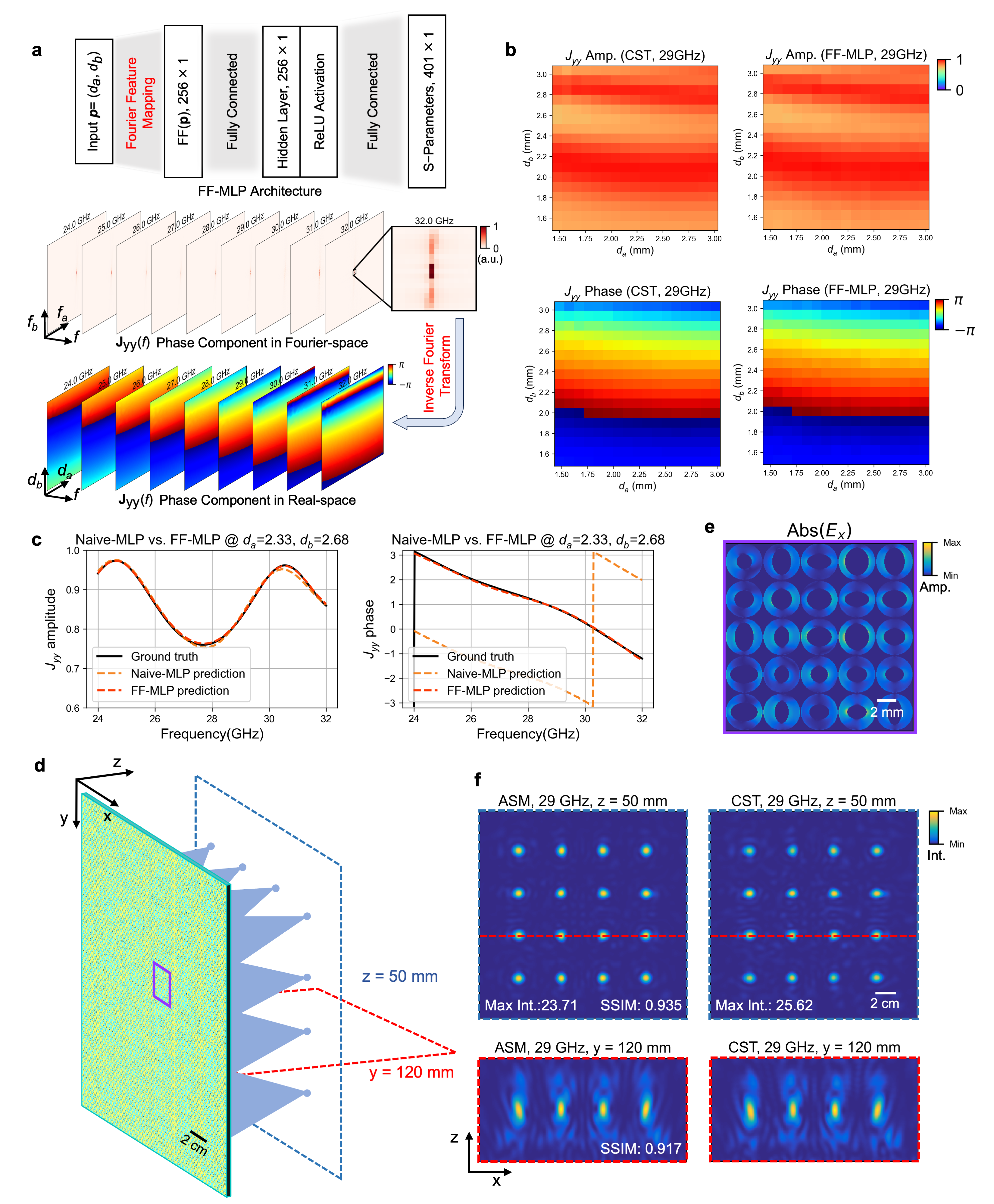}
  \captionsetup{font={small,stretch=0.85}}
  \caption{\textbf{Mini-metanets for predicting EM field response of meta-atoms.} \textbf{a} Mini-metanets are constructed using the FF-MLP architecture (top) to predict the $S_{21}$ parameters of meta-atoms by mapping into the Fourier-space. Using the predicted $S_{21}$ parameters, the Jones matrix can be obtained. The phase component of matrix element $\mathbf{J}_{yy}(\emph{f})$ in Fourier-space (middle) and real-space (bottom) are shown in the figure. \textbf{b} Comparisons between CST and FF-MLP prediction for the amplitude and phase of $\mathbf{J}_{yy}(\emph{f})$ under different design parameters at 29~GHz demonstrate the effectiveness of FF-MLP in characterizing the EM response of meta-atoms. \textbf{c} At specific design parameters ($d_a$ = 2.33~mm, $d_b$ = 2.68~mm), the frequency-dependent variation of $\mathbf{J}_{yy}(\emph{f})$ amplitude (left) and phase (right) is shown. \textbf{d, e, f} The internal resonance characteristics of meta-atoms are evaluated by examining the amplitude response at x-polarization, and the high-accuracy modeling of EM field modulation and propagation can be verified by comparing the optical field between our model and CST, with the exemplar task of multi-focusing.}
  \label{fig2}
\end{figure}

Meta-atom represents the basic unit of the DMNN, the high-dimensional EM response of which can be precisely characterized with mini-metanets by utilizing its internal resonance characteristics. To predict the multi-frequency Jones matrix, we train four mini-metanets to generate the mapping between input geometric parameters $d_a$ and $d_b$, i.e., elliptical major and minor axes, respectively, and four output $S_{21}$ parameters, i.e., the output multi-spectral amplitude and phase modulations at the x- and y-polarizations. Fig.~\ref{fig2}\textbf{a} shows the mini-metanet (top) and exemplar phase component of $\mathbf{J}_{yy}(\emph{f})$ in Fourier-space (middle) and real-space (bottom). The mini-metanet is a FF-MLP architecture, which first maps the input geometric parameters to a 256-dimensional feature space via Fourier feature encoding. Then, the fully-connected hidden layer and output layer are utilized to produce a 401-dimensional response vector that spans frequencies from 24~GHz to 32~GHz with a resolution of 0.02~GHz. Besides, the geometric parameter $\theta$ in the rotation matrices $\mathbf{R}(\theta)$ and $\mathbf{R}(-\theta)$ introduces the polarization conversion that changes the polarization direction and the Jones vector. When $\theta = 0^\circ$, the structure does not induce polarization conversion, and thus the corresponding Jones matrix reduces to a diagonal form. Under this condition, we characterize the response using four quantities: $\mathbf{A}_{x}(\emph{f})$ and $\mathbf{\Phi}_{x}(\emph{f})$ for the amplitude and phase of x-polarized transmission under x-polarized incidence, and $\mathbf{A}_{y}(\emph{f})$, $\mathbf{\Phi}_{y}(\emph{f})$ for the y-polarized counterpart. When $\theta \ne 0^\circ$, the complete Jones matrix can be reconstructed by applying a rotation matrix, as detailed in Supplementary Note~1.

To train the mini-metanet, we first performed full-wave simulations using CST Microwave Studio to characterize the modulation and generate the training and test datasets by setting $\theta = 0^\circ$ (see Methods), with which the frequency-dependent variations of the four response characteristics as functions of $d_a$ and $d_b$ are modeled. This enables rapid and accurate prediction of the frequency-dependent Jones matrices, facilitating physically characterize the HDM of meta-atom through the data-driven inference. To efficiently generate the training dataset, we discretized $d_a$ and $d_b$ within the range 1.4~mm to 3.1~mm with a coarse step size of 0.05~mm, resulting in 1025 unique metasurface configurations. Full-wave simulations were performed in CST to extract the EM responses of these configurations. To enhance the training data without incurring additional simulation cost, we employed the data augmentation strategy by interpolating in the parameter space at a finer resolution of 0.01~mm, thereby expanding the dataset by a factor of 25. After training, the FF-MLPs provide a highly efficient and differentiable mapping from structural design parameters to EM response functions that facilitates the meta-training process for different tasks.

To test the generalization of trained models for predicting EM responses, we acquired an independent set of EM responses by sampling $d_a$ in the range of 1.43~mm to 3.03~mm and $d_b$ from 1.48~mm to 3.08~mm, with a step size of 0.1~mm for both parameters. This resulted in 289 distinct metasurface configurations, which were used exclusively for performance evaluations. We quantitatively evaluate the prediction results of different Jones matrix elements by varying $d_a$, $d_b$, and $\emph{f}$. Fig.~\ref{fig2}\textbf{b} and \textbf{c} present the predicted results for $\mathbf{J}_{yy}(\emph{f})$ amplitude and phase components, compared with CST characterizations as the ground truth, at 29~GHz with varying $d_a$ and $d_b$, and across different frequencies with a fixed design point $d_a = 2.33~\mathrm{mm}$ and $d_b = 2.68~\mathrm{mm}$. As a baseline for comparison, we also implemented a naive-MLP with two hidden layers, each comprising 256 neurons and having the same number of network parameters. We further compare the prediction errors of two models. The FF-MLP achieves amplitude prediction errors consistently below 0.01 and phase errors below 0.1~rad, with a phase mean absolute error of 0.024~rad. In contrast, the naive-MLP yields a higher phase mean absolute error of 0.088~rad. Detailed prediction results across frequencies are provided in Supplementary Note 1. The close alignment between the interpolated and ground-truth results demonstrates the effectiveness of the prediction method. Notably, the naive-MLP exhibits significant prediction failures at certain design parameters, underscoring the robustness of the FF-MLP approach.

To evaluate the coupling between meta-atoms in the designed metasurface, we construct a multi-focal system, considering the full-wave simulation efficiency, with 16 x-polarization focusing spots on the detection regions at the focusing distance of 50~mm by using a metasurface with the same number of meta-atoms, i.e., 50$\times$50, as shown in Fig.~\ref{fig2}\textbf{d}. Due to the surrounding metallic frame, the resulting resonant fields, presented with the amplitude of x-polarization, are primarily confined within the region between the ellipse and the surrounding frame, indicating strong EM localization and reduced coupling across adjacent units (see Fig.~\ref{fig2}\textbf{e}). Based on the mini-metanet predictions at each meta-atom, Fig.~\ref{fig2}\textbf{f} further compares the EM fields of the forward model with those obtained from CST on the x-y detection plane and x-z propagation plane. To quantitatively assess the similarity of the EM fields between our model and CST, the structural similarity index measure (SSIM) of energy distribution is calculated, which is 0.935 for the energy distribution on the x-y detection plane and 0.917 for a specific x-z propagation plane at $y = 120$~mm. The high similarities of spatial energy distributions validate the accuracy of our model.

\subsubsection*{Numerical evaluation of all-optical DMNN}

The all-optical DMNN architecture is designed to perform high-precision DOA estimation with high-throughput. The incident optical field corresponding to a specific incident direction, polarization state, and frequency can be modeled as:
\begin{equation}
\mathbf{E}(x, y; \lambda) = \vec{p} E_0 e^{j \frac{2\pi}{\lambda} (x \sin\alpha \cos\beta + y \sin\alpha \sin\beta)},
\end{equation}
where $\vec{p}$ denotes the polarization vector, $E_0$ is the amplitude, and $\lambda$ is the wavelength. When $\beta = 0^\circ$, $\alpha$ corresponds to the azimuthal angle; whereas for $\beta = 90^\circ$, $\alpha$ corresponds to the elevational angle. The polarization multiplexing capability enables the DMNN to simultaneously estimate azimuth and elevation angles by utilizing x- and y-polarization channels of incident light, respectively. Given that the target emits EM waves spanning multiple frequencies and polarization states, the DMNN can process these high-dimensional signals in parallel at the speed of light. For each frequency polarization pair, the forward propagation is independently modeled using the angular spectrum method (ASM), while the modulation behavior of the metasurface is characterized by trained FF-MLP models (see Supplementary Note 2 for details). This hybrid modeling framework allows for efficient simulation of the forward process of the DMNN across all dimensions of the input space. Implementation details are provided in the Methods section. By optimizing the design parameters of each metasurface unit through supervised training, the network learns to direct incident optical energy from different directions to corresponding detector regions in the detection plane, with which the estimation accuracy of the model can be obtained. By analyzing the energy distributions across the detector regions, i.e., top two detectors with largest and the second-largest intensities in this work, in the x- and y-polarized output channels, the network estimates either the azimuthal or the elevational angle of the target. By integrating the estimation results across multiple frequencies with interleaved angular intervals, a more accurate and robust angular estimation can be achieved, analogous to the concept of ensemble learning in machine learning (see Supplementary Note 3).

To enhance the performance of DMNN, we proposed a stage-wise training strategy, which can be divided into two stages (detailed in Supplementary Note 4 and Methods). In the first stage, the output energy distribution is used as the ground truth to guide the concentration of the output field, with a relatively large learning rate employed to accelerate convergence. In the second stage, the training objective shifts to the detection regions by using the target one-hot vector as the ground truth, with a reduced learning rate used to fine-tune the network parameters. As shown in Fig.~\ref{fig3}\textbf{a}, \textbf{b}, and \textbf{c}, we demonstrate the effectiveness of the proposed stage-wise training method by designing the DMNN with 16 detection regions, each corresponding to an angular interval of $1.5^\circ$. Three operational frequencies, i.e., 27~GHz, 29~GHz, and 31~GHz, were selected for evaluations. The corresponding FOVs are defined as [$-12.5^\circ$, $11.5^\circ$], [$-12^\circ$, $12^\circ$], and [$-11.5^\circ$, $12.5^\circ$], respectively. Fig.~\ref{fig3}\textbf{d} presents a comparative analysis of the simulation performance before and after applying the stage-wise training approach. Without this strategy, the average task accuracy across all frequency and polarization dimensions is $80.3\%$. After adopting the two-stage training to generate the super-oscillatory angular response within the FOV, the accuracy improves significantly to $99.0\%$, clearly demonstrating the effectiveness of the proposed method in optimizing multi-dimensional DOA estimation.

The azimuthal and elevational angles are estimated by utilizing the top two energy detection regions (see Supplementary Note 3 for details). Numerical evaluation shows that our method achieves an estimation mean error of $0.07^\circ$ for the single target and $0.17^\circ$ for two incoherent targets within a FOV of [$-11.5^\circ$, $11.5^\circ$] (see Fig.~\ref{fig3}, Supplementary Fig.~8 and 9). To quantitatively evaluate the computing performance and throughput of DMNN for DOA estimation, we further define the metric termed angular estimation throughput ($AET$), which is formulated as follows:
\begin{equation}
AET = \frac{N_d \times N_t \times FOV}{\text{Mean Error}},
\end{equation}
where $N_d$ denotes the number of resolvable angular directions, $N_t$ denotes the number of targets, and $FOV$ is the FOV in degrees. This definition provides a balanced and intuitive metric that jointly accounts for both the angular resolution or estimation accuracy and the coverage range. In conventional designs that formulate the task with discrete classification strategy \cite{gao2024superresolution}, each detection region corresponds to a fixed angular interval. Improving the AET in such cases often requires deploying multiple physical systems via spatial multiplexing or implementing multiple network configurations using time-division multiplexing. For comparison, a conventional configuration with 16 discrete angular intervals under a single frequency and single polarization yields a theoretical upper-bound AET of 64, assuming all correct classification performance and using midpoint as the reference. Our approach achieves an AET of 657.1 for the single target, which is over an order of magnitude higher than the conventional method, and 541.1 for two incoherent targets with random angular separation. Furthermore, we validated the scalability of DMNN across larger angular ranges. Specifically, for a FOV of [$-23.0^\circ$, $23.0^\circ$], we obtained an average estimation error of $0.13^\circ$, resulting in an AET of 707.7. When the FOV was further expanded to [$-46.0^\circ$, $46.0^\circ$], the average estimation error increased to $0.36^\circ$, corresponding to an AET of 511.1 (see detailed results in Supplementary Note 4).

\subsubsection*{Experimental validation of all-optical DMNN}

\begin{figure}
\centering
  \includegraphics[width=0.9\linewidth]{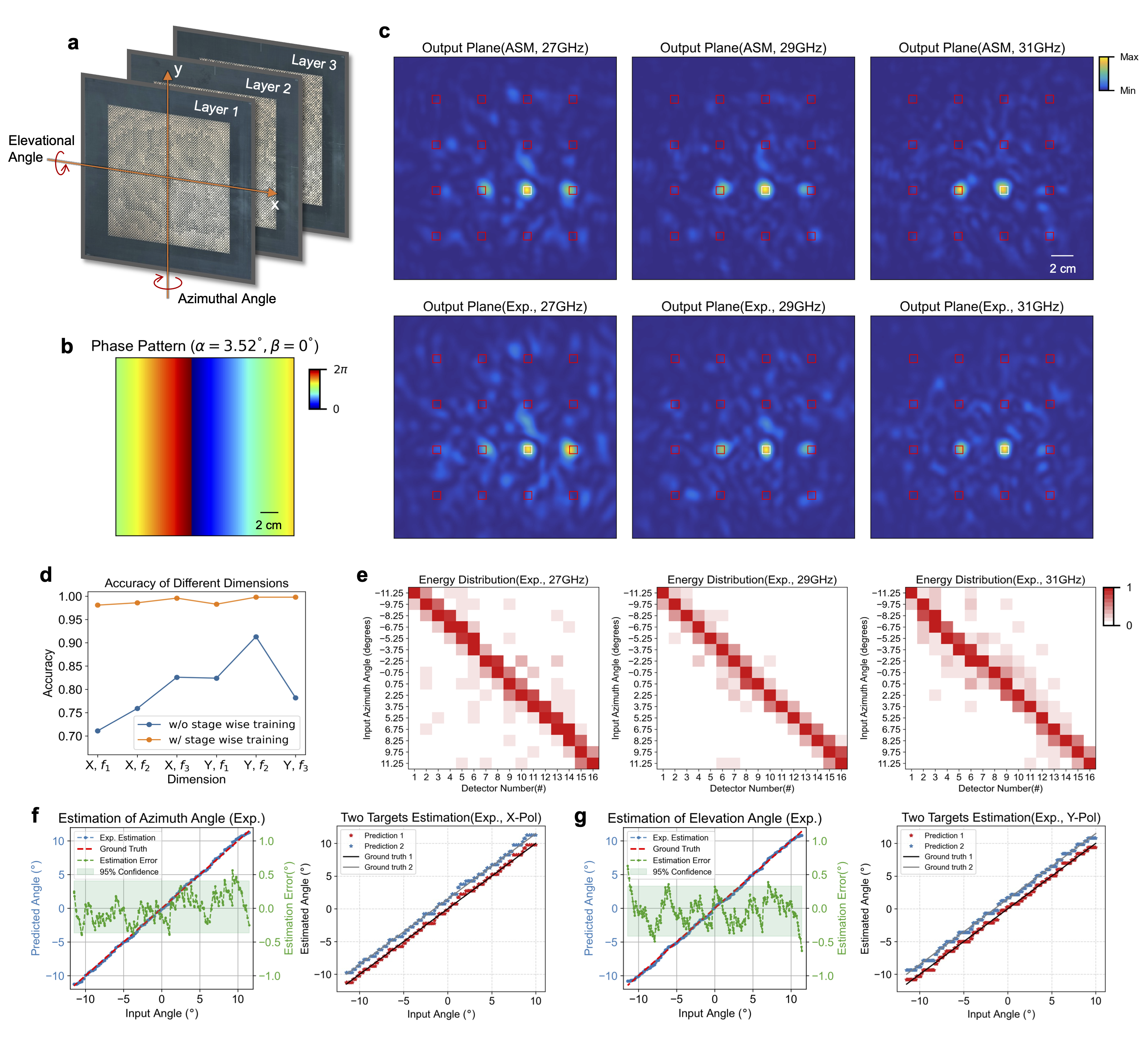}
  \caption{\textbf{All-optical super-resolution angle estimation with high-throughput.} \textbf{a} Three-layer DMNN designed with meta-training approach, which has 16 detection regions and $1.5^\circ$ angular interval, multiplexing three frequencies (27, 29 and 31~GHz) and 2 polarizations (x- and y-polarization). \textbf{b} X-polarized phase distribution of incidence light with an azimuthal angle of $3.52^\circ$ for estimation. \textbf{c} Output intensity distribution of our forward model (top) and experimental results (bottom) at 27, 29, and 31~GHz. \textbf{d} The improvement of accuracy, i.e., the confidence value of DOA estimation, at different multiplexing dimensions using the proposed stage-wise training method. \textbf{e} Experimental energy distributions of 16 detection regions under 16 incident angles at 27, 29, and 31~GHz, with the angles chosen as the midpoints for the 29~GHz case. \textbf{f, g} Experimental results of DOA estimations with HDM for the single target and two incoherent targets separated by $1.5^\circ$ under different azimuthal and elevational angles.}
  \label{fig3}
\end{figure}

For experimental validation, the designed three-layer DMNN has a $50 \times 50$ array of trainable meta-atoms with $200~\mathrm{mm} \times 200~\mathrm{mm}$ effective modulation aperture size at each layer (see Methods). Therefore, the diffraction-limited angular resolution, defined as $1.22\lambda/\phi$, with $\phi = 200~\mathrm{mm}$ being the aperture size, is $3.88^\circ$, $3.61^\circ$, and $3.38^\circ$ at the operation frequencies of 27 GHz, 29 GHz, and 31 GHz, respectively. To facilitate the experiments, we separately estimate the azimuthal and elevational angles of target antennas within the FOV. We first evaluated the system's azimuthal estimation capability using x-polarized incident waves. A representative input with an azimuthal angle of $3.52^\circ$ was selected, with the corresponding input phase pattern shown in Fig.~\ref{fig3}\textbf{b}. The numerically and experimentally measured energy fields at 27, 29, and 31~GHz are shown in Fig.~\ref{fig3}\textbf{c}. All three frequencies exhibit peak energy concentration at the 11-th detection region, indicating consistent directional sensitivity. Energy leakage into non-detection areas was observed, particularly due to the second training stage, which improves classification accuracy at the cost of slight field energy perturbation.

Fig.~\ref{fig3}\textbf{e} presents the normalized energy distributions over the 16 detection regions for all three frequencies. These correspond to the angular midpoints calibrated at 29~GHz. At this frequency, the energy distributions are relatively concentrated, while slight lateral shifts are observed at 27 and 31~GHz due to their deviation from the center of their angular coverage. Angle estimation is performed by identifying the top two detection regions with maximum and second-maximum values in the energy distributions across the three frequencies. By statistically analyzing the estimated results under different incident angles (see Methods), the experiment achieved an average estimation error of $0.16^\circ$ within the FOV of [$-11.5^\circ$, $11.5^\circ$], as shown in Fig.~\ref{fig3}\textbf{f}(left). The green curve shows the estimation error for each angle, and the green region indicates the error range of [$-0.36^\circ$, $0.41^\circ$], corresponding to a $95\%$ confidence interval for the angle estimation. For elevational angle estimation under y-polarization (see Fig.~\ref{fig3}\textbf{g}(left)), the results achieved an average estimation error of $0.15^\circ$, thus leading to an experimental AET of 287.5.

To evaluate the capability of all-optical DMNN in estimating multiple target sources from different directions, we take the azimuthal and elevational angle estimations for two incident sources separated by $1.5^\circ$ that is beyond the diffraction limit as a representative example. When the two sources are located in distinct angular intervals, their directions can be estimated by identifying the two largest responses among the 16 detection regions. In practical scenarios, we assume that two sources are incoherent, such that the resulting output energy is the linear superposition of the output energy produced by each source (see more details in Supplementary Note 5). Based on the pre-characterized single source energy responses collected during the initial phase, we synthesize the two source energy distributions under this assumption (see Methods). Fig.~\ref{fig3}\textbf{f}(right) and Fig.~\ref{fig3}\textbf{g}(right) show the experimental azimuthal and elevational angle estimation results for two targets separated by $1.5^\circ$, respectively, achieving an average estimation error of $0.23^\circ$ and $0.24^\circ$, respectively. The experimental results validate the effectiveness of DMNN for super-resolution and high-throughput DOA estimation at high-accuracy based on the HDM and super-oscillatory angular response characteristic.

\subsubsection*{Super-resolution angle estimation with optoelectronic DMNN}

\begin{figure}
\centering
  \includegraphics[width=0.9\linewidth]{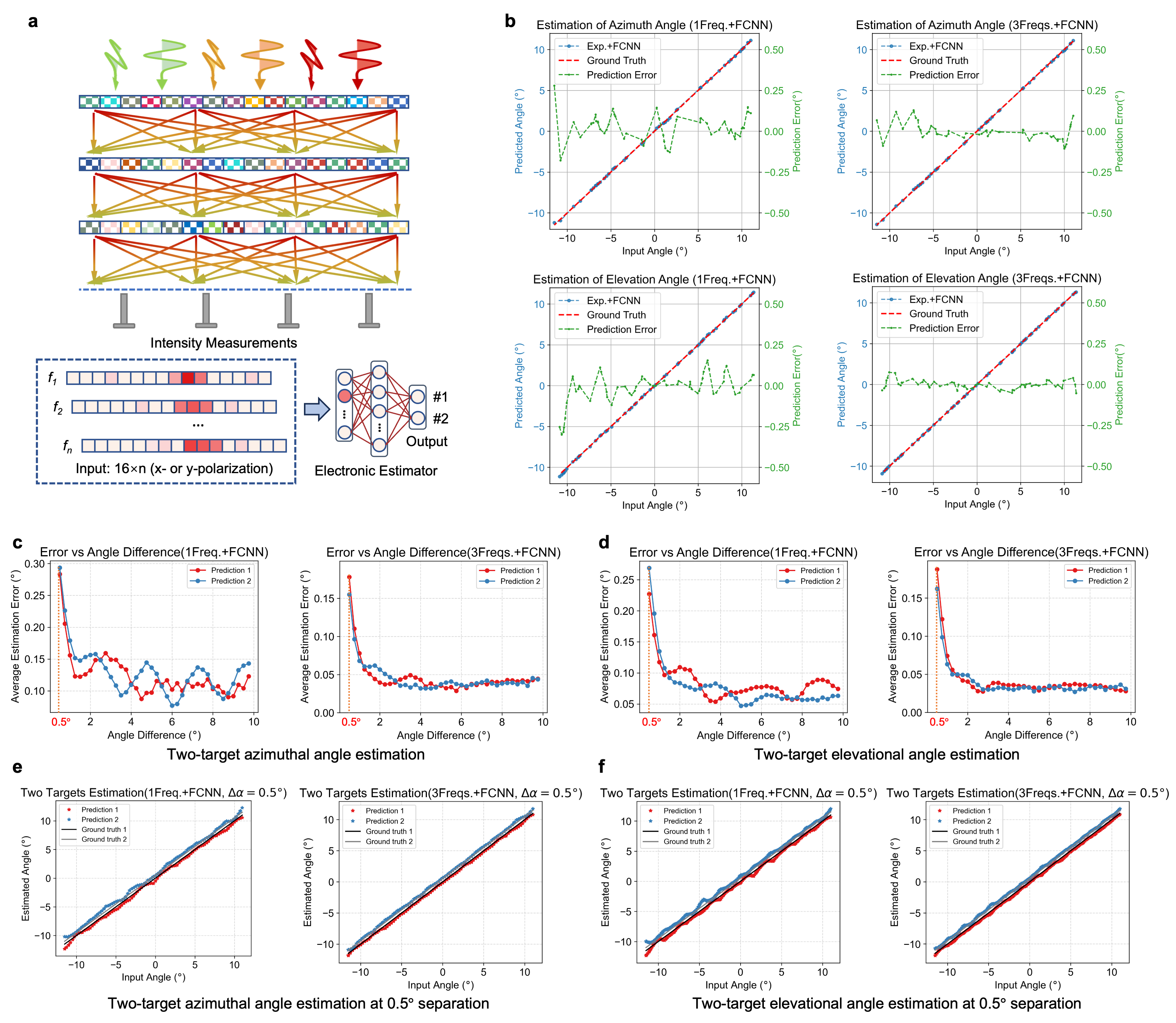}
  \caption{\textbf{Optoelectronic DMNN for super-resolution DOA estimation with high-performance.} \textbf{a} Incorporating the lightweight electronic neural network as the angular estimation post-processor to extract the target angles from intensity measurements, where the output node number corresponds to the target number. \textbf{b} Experimental validations of the single-target azimuthal (top) and elevational (bottom) angle estimations using the detected energy distribution of a single frequency and three frequencies. \textbf{c, d} Experimental results of azimuthal and elevational angle estimation, quantitatively evaluated with mean errors, for two targets separated by different angles based on the detected energy distribution of a single frequency (left) and three frequencies (right). \textbf{e, f} Experimental results of DOA estimations for two incoherent targets separated by $0.5^\circ$ under different azimuthal and elevational angles based on the detected energy distribution of a single frequency (left) and three frequencies (right).}
  \label{fig4}
\end{figure}

To improve the angular estimation resolution and network performance, we propose to design the optoelectronic DMNN that integrates the DMNN with a electronic estimator. This collaborative architecture allows to fully exploit the multi-dimensional optical information captured at the detection plane. As illustrated in Fig.~\ref{fig4}\textbf{a}, the DMNN with super-oscillatory angular responses serves as the all-optical modulation front-end, processing the incident EM field and focusing energy into 16 spatially distributed detection regions, thereby significantly reducing the dimensionality of the optical data. Subsequently, an electronic estimator is employed to analyze the energy distributions across different frequencies and polarizations. This estimator can be customized for either single-target or multi-target angle estimation based on the number of output nodes, allowing for flexible application across diverse sensing scenarios.

We first validated the performance of the proposed optoelectronic system for the azimuthal and elevational angle estimation of a single target (see Fig.~\ref{fig4}\textbf{b}). In this validation, we randomly sample the incident angles ranging from $-11.5^\circ$ to $11.5^\circ$, each corresponding to the energy distributions of x- or y-polarized waves at 27, 29, and 31~GHz. To assess the benefit of multi-dimensional information fusion, we compared the estimation performance using only the energy distribution at 29~GHz with that using all three frequencies, as shown in Fig.~\ref{fig4}\textbf{b}(left) and Fig.~\ref{fig4}\textbf{b}(right), respectively. As the DMNN has already compressed the optical information into 16 detection regions, we employed a lightweight fully connected neural network with a single hidden layer to perform the final angle regression. Using only single-frequency input resulted in an estimation mean error of $0.066^\circ$ and $0.063^\circ$ for azimuthal and elevational angles, respectively, while incorporating multi-frequency energy features with optical coding scheme reduced the mean error to $0.037^\circ$ and $0.019^\circ$, respectively, demonstrating the advantage of spectral diversity in enhancing angular resolution.

We further evaluated the performance of the proposed optoelectronic system in estimating the azimuthal and elevational angles for two sources, with the estimation mean error versus different angle separation shown in Fig.~\ref{fig4}(\textbf{c, d}). Based on the pre-collected single-target energy responses, synthetic two source energy distributions were generated by linearly superimposing the output energy of two independent sources. Similar to the single-target case, we compared the estimation performance using energy distributions from a single frequency (29~GHz) versus those incorporating all three frequencies (27, 29, and 31~GHz). The average azimuthal angle estimation errors of two targets at different input angular separation were $0.135^\circ$ and $0.048^\circ$ for single frequency and three frequencies, respectively. Besides, the average elevational angle estimation errors of two targets at different input angular separation were $0.106^\circ$ and $0.047^\circ$ for single frequency and three frequencies, respectively. The corresponding AET, averaged on the azimuthal and elevational angle estimation, improved from 763.5 to 1916.7 with the multi-frequency coding. The results indicate the significant improvements in accuracy through multi-frequency coding scheme with a super-resolved angular resolution of $0.5^\circ$, which is $7\times$ higher than the Rayleigh diffraction limits.

To illustrate the estimation behavior under different source angular separations, Fig.~\ref{fig4}(\textbf{e, f}), Supplementary Fig.~12 and 13 show the representative results for azimuthal and elevational angular separations of $0.5^\circ$, $1.5^\circ$, and $3.0^\circ$, by using a single-frequency and multi-frequency energy features. In these figures, red and blue solid dots represent the predicted azimuthal or elevational angles of the two sources, corresponding to the smaller and larger angles, respectively. When multiple frequencies were employed, the mean absolute errors were $0.17^\circ$, $0.05^\circ$, and $0.04^\circ$ for the $0.5^\circ$, $1.5^\circ$, and $3.0^\circ$ cases, respectively; in contrast, using single-frequency features yielded larger errors of $0.27^\circ$, $0.12^\circ$, and $0.11^\circ$. As evident from these comparisons, incorporating multiple frequencies significantly improves the estimation accuracy, especially in challenging cases with small angular separation.

\subsection*{Discussion}

Based on the high-dimensional EM multiplexing and coding scheme with the super-oscillatory angular response characteristic~\cite{chen2019superoscillation}, the three-layer DMNN achieves the angular resolution of $0.5^\circ$ for two target estimation that is $\sim$7$\times$ higher than the Rayleigh diffraction-limited angular resolution defined by the system aperture size. In addition, with the metasurface layers for spatial frequency modulations, the generated super-oscillatory frequencies for angular resolution also surpass the latent spatial frequency upper-bound caused by the free-space diffractive propagation, demonstrating the advanced super-resolution sensing capabilities of the DMNN compared with conventional sensing systems.

% In our current experimental implementation, multidimensional measurements were performed sequentially by switching the vector network analyzer (VNA) between different frequency settings and mechanically reconfiguring the waveguide probe for various polarization states. This sequential acquisition inherently limits the overall measurement speed, as each dimension requires an independent measurement cycle. Looking forward, the capability to perform simultaneous multidimensional acquisition, such as recording multiple frequency components using a spectrum analyzer together with various polarization states in a single measurement cycle, would significantly enhance the efficiency of the system. Furthermore, incorporating RF switches to rapidly select different measurement regions without mechanical repositioning could further reduce acquisition time. This improvement would better exploit the intrinsic advantages of multidimensional measurement, including richer feature extraction, improved robustness against temporal instability, and the potential to capture transient or dynamic phenomena that are inaccessible with slow sequential scans.

In our current experimental implementation, multidimensional measurements were performed sequentially by switching the vector network analyzer (VNA) between different frequency settings and mechanically reconfiguring the waveguide probe for various polarization states. This sequential acquisition inherently limits the overall measurement speed, as each dimension requires an independent measurement cycle. Looking forward, increasing the number of polarization channels could further enhance angular estimation coverage and accuracy, e.g., inferring both $\alpha$ and $\beta$ values. And the capability to perform simultaneous multidimensional acquisition, such as recording multiple frequency components using a spectrum analyzer together with various polarization states in a single measurement cycle, would significantly enhance the efficiency of the system. Furthermore, incorporating RF switches to rapidly select different measurement regions without mechanical repositioning could further reduce acquisition time. This improvement would better exploit the intrinsic advantages of multidimensional measurement, including richer feature extraction, improved robustness against temporal instability, and the potential to capture transient or dynamic phenomena that are inaccessible with slow sequential scans.

The proposed architecture demonstrates strong scalability potential, primarily owing to the lightweight nature of the mini-metanet modules. This design enables seamless integration into large-scale DMNNs for end-to-end optimization. While the current metasurface implementation offers promising functionality, its relatively limited design degrees of freedom may pose challenges in independently modulating multiple dimensions of the electromagnetic field. To address this, an effective method is to increase the number of modulation units, enabling global optimization and reducing task interference across different dimensions. Another complementary strategy involves enhancing the design flexibility of individual units to produce more diverse and programmable modulation responses. These directions collectively offer promising pathways to further elevate the adaptability and performance of DMNN-based systems in complex electromagnetic environments.

In summary, we have proposed DMNN architecture and have experimentally validated its application in all-optical super-resolution DOA estimation with high-throughout. By utilizing multi-dimensional EM field information, we can enhance the angle estimation throughput and improve estimation resolution under the same FOV. The proposed deep learning metasurface based on FF-MLP enhances the prediction stability of HDM metasurfaces through the novel introduction of Fourier feature operators. Furthermore, we experimentally demonstrated that using DMNN as the photonic computational front end and a lightweight electronic neural network as the back end further improves angle estimation accuracy, from $0.15^\circ$ to $0.028^\circ$. In future, DMNNs are expected to take on the capability of rapid information processing of high-dimensional EM fields, with potential for widespread application in various fields such as communications, radar, imaging, and sensing.

%%%%%%%%%%%%%%%% MATERIALS AND METHODS %%%%%%%%%%%%%%%
\subsection*{Materials and Methods}
\subsubsection*{DMNN configuration and fabrication}

The designed DMNN system for super-resolution DOA estimation consists of three modulation layers, each composed of a $50 \times 50$ array of meta-units, forming an effective modulation area of $200~\mathrm{mm} \times 200~\mathrm{mm}$. To ensure mechanical stability and facilitate mounting, an additional dielectric margin made of F4BM245 substrate is included around the modulation region, yielding a total board size of $274.5~\mathrm{mm} \times 274.5~\mathrm{mm}$. The interlayer spacing between adjacent metasurface layers and the distance from the last layer to the detection plane are both fixed at $100~\mathrm{mm}$. In terms of architectural choice, we found that additional layers may improve performance but only marginally, while incurring substantially higher fabrication cost and simulation overhead. Therefore, a three-layer configuration was adopted as a balanced compromise between achievable accuracy and practical feasibility. The angle estimation system operates at three discrete frequencies: 27~GHz, 29~GHz, and 31~GHz.

The metasurfaces are fabricated using standard printed circuit board (PCB) process with the dielectric material F4BM245, which has a relative permittivity of approximately 2.60 and a loss tangent of 0.003 from 24~GHz to 32~GHz. To broaden the operational bandwidth and achieve $2\pi$ phase modulation, a multi-layer structure is implemented. The meta-atom size is set to $p = 4$~mm, and the substrate thickness for each layer is $h = 0.7$~mm, a total of 5 layers. The copper layer thickness is $t = 0.035$~mm, with 6 layers in total. During fabrication of meta-atoms, the elliptical meta-atoms are designed with tunable semi-major and semi-minor axes ranging from $1.4~\mathrm{mm}$ to $3.1~\mathrm{mm}$ with a manufacturing tolerance within $\pm 0.025~mm$, ensuring full $2\pi$ phase coverage and average phase error induced by fabrication below $\pm 5^\circ$.

\subsubsection*{Experimental setup and data sampling}

The angle estimation experiments were carried out in a microwave anechoic chamber (Supplementary Fig.~10) using a Keysight P5006B vector network analyzer (VNA). Port~1 of the VNA was connected to a transmitting horn antenna and Port~2 to a waveguide probe, with the energy distribution characterized from the measured transmission coefficient ($S_{21}$). The horn antenna was positioned approximately $7~\mathrm{m}$ from the DMNN, ensuring plane-wave incidence. The DMNN and the probe were mounted on a motorized XY stage with angular and translational resolutions of 0.01$^\circ$ and $0.01~\mathrm{mm}$, respectively. By rotating the mounting stage, incident waves at different angles were emulated, while the probe translation enabled sampling of spatial energy distributions at various positions.

For each azimuthal or elevational angle estimation task, approximately 260 angular samples were collected within the range of $-11.5^\circ$ to $11.5^\circ$, under both x- and y-polarized incidence at 27, 29, and 31~GHz. In the all-optical angle estimation task, all single-target samples were directly used for testing. In the optoelectronic angle estimation task, $80\%$ of the data were used for training and $20\%$ for testing. Two-target datasets were synthetically generated by incoherent superposition of single-target measurements, yielding 3000 randomly generated samples, again divided into 80\% training and 20\% testing. To further assess estimation errors for prescribed angular separations, we constructed exhaustive two-target test cases, including configurations excluded from training.

\subsubsection*{Training of electronic neural network estimator}

To achieve higher-accuracy angle estimation, a lightweight electronic post-processing neural network with a single hidden layer was employed to further extract energy features from different detection regions. For the single-target angle estimation task, the hidden layer consisted of 64 neurons activated by the ReLU function, and the network directly output the predicted incidence angle. For the dual-target incoherent angle estimation task, the hidden layer comprised 256 ReLU-activated neurons, and the output layer produced the estimated angles of both target. The batch size was fixed at 16, and the Adam optimization algorithm was adopted. The learning rates were set to 0.005 and 0.001 for the single-source and dual-source tasks, respectively, with training epochs of 150 and 300.

\subsubsection*{Stage-wise training of DMNN}

To optimize the performance of the DMNN, we adopt a stage-wise training strategy consisting of two consecutive stages, each employing distinct loss functions and learning rate schedules. Let $\{\mathbf{D}^{(l)}_{i,j} \in \mathbb{R}^{3} \mid i,j = 1,2,\dots, N;\ l = 1,2,\dots, L\}$ denote the design parameter vectors for each modulation unit $(i,j)$ at layer $l$. Each vector $\mathbf{D}^{(l)}_{i,j}$ is mapped via pre-trained FF-MLPs to a frequency-dependent Jones matrix:
\begin{equation}
\mathbf{J}^{(l)}_{{\lambda}}(x, y) = \mathbf{M}_{{\lambda}}\left( \mathbf{D}^{(l)}_{i,j} \right), \quad 
i = \left\lfloor \frac{x}{a} \right\rfloor + 1,\quad 
j = \left\lfloor \frac{y}{a} \right\rfloor + 1,
\end{equation}
where $(x, y)$ are spatial coordinates and $a$ denotes the unit size of the modulation element. For a given high-dimensional EM field input $\mathbf{E}_{\lambda}^{\text{in}}(x,y)$, through the parallel inference of DMNN (see Supplementary Note 2 and 4 for details), the output EM field at the detection plane is computed as:
\begin{equation}
\mathbf{E}_{\lambda}^{\text{det}}(\mathbf{D}) = \mathbf{E}_{\lambda,\text{in}}^{(L+1)} = \prod_{l=1}^L {\mathbf{U}_{\lambda}\mathbf{M}_{\lambda}(\mathbf{D}^{(l)})}\mathbf{E}_{\lambda}^{\text{in}}   
\end{equation}
\noindent where spatial coordinates $(x, y)$ are omitted for clarity. For a certain linear polarization detection $\boldsymbol{p} = [\cos\theta,\ \sin\theta]$, the corresponding detected intensity becomes:
\begin{equation}
I_{{\lambda}, \boldsymbol{p}}(\mathbf{D}) = \left| \boldsymbol{p} \cdot \mathbf{E}^{\mathrm{det}}_{{\lambda}}(\mathbf{D}) \right|^2,
\end{equation}

In the first stage, the training objective is to align the predicted intensity distribution with the target intensity pattern $\mathrm{TI}_{{\lambda}, \boldsymbol{p}}$ at the output plane for each dimension and polarization:
\begin{equation}
\min_{\mathbf{D}} \left( \sum_{{\lambda}} \sum_{\boldsymbol{p}} \epsilon_{{\lambda}, \boldsymbol{p}}^{(1)} \cdot \mathrm{MSE}\left( I_{{\lambda}, \boldsymbol{p}}(\mathbf{D}),\ \zeta_{{\lambda}, \boldsymbol{p}}^{(1)} \cdot \mathrm{TI}_{{\lambda}, \boldsymbol{p}} \right) \right),
\end{equation}
where $\epsilon_{\boldsymbol{\lambda}, \boldsymbol{p}}^{(1)}$ and $\zeta_{{\lambda}, \boldsymbol{p}}^{(1)}$ are balancing coefficients. A relatively large learning rate ($\mathrm{lr} = 0.05$) is used to accelerate convergence. The training set consists of 4,800 samples per dimension, uniformly distributed across 16 classes (300 samples each), with batch size 100.

In the second stage, the focus shifts to improving accuracy within specific detection regions. The intensity field is integrated over 16 predefined detector regions, forming a regional energy distribution $P_{{\lambda}, \boldsymbol{p}}(\mathbf{D}) \in \mathbb{R}^{16 \times 1}$. The optimization goal becomes:
\begin{equation}
\min_{\mathbf{D}} \left( \sum_{{\lambda}} \sum_{\boldsymbol{p}} \epsilon_{{\lambda}, \boldsymbol{p}}^{(2)} \cdot \mathrm{MSE}\left( P_{{\lambda}, \boldsymbol{p}}(\mathbf{D}),\ \zeta_{{\lambda}, \boldsymbol{p}}^{(2)} \cdot \mathrm{TD}_{{\lambda}, \boldsymbol{p}} \right) \right),
\end{equation}
where $\epsilon_{\boldsymbol{\lambda}, \boldsymbol{p}}^{(2)}$ and $\zeta_{{\lambda}, \boldsymbol{p}}^{(2)}$ are balancing coefficients as well, $\mathrm{TD}_{{\lambda}, \boldsymbol{p}}$ denotes the target energy distribution across detectors. A smaller learning rate ($\mathrm{lr} = 0.02$) is employed to enhance stability and prevent energy over-suppression. Although the overall field distribution may become slightly more dispersed across the detection plane, the focused energy within the 16 target regions leads to a significant improvement in task-specific accuracy.

%%%%%%%%%%%%%%%% REFERENCES %%%%%%%%%%%%%%%

\clearpage % Clear all remaining figures and tables then start a new page

% The list of references goes after the main text and before the acknowledgements
% When preparing an initial submission, we recommend you use BibTeX, like this:
%
% \bibliography{reference} % for a file named science_template.bib

\bibliographystyle{sciencemag}

% After the paper has completed peer review and been revised ready for acceptance,
% you should comment out the lines above and copy-paste the contents of your .bbl
% file here instead. This will help ensure that our conversion software works correctly.
% Remember to re-run BibTeX first - check the timestamp!
%
% Example of the first three entries copy-pasted from science_template.bbl:
%
%\begin{thebibliography}{1}
%
%\bibitem{example}
%A.~N. {Author}, An example reference. \emph{Journal of Improbable Research}
%  \textbf{1}, 67 (2020).
%
%\bibitem{example2}
%F.~M. {Surname}, S.~{Author}, A second example. \emph{Interesting Research
%  Letters} \textbf{32}, 897 (2019).
%
%\bibitem{example_preprint}
%P.~{One}, P.~{Two}, P.~{Three}, {An unpublished preprint}. \emph{preprint}
%  (2021), arXiv:2101.12345.
%
%\end{thebibliography}

%%%%%%%%%%%%%%%% ACKNOWLEDGEMENTS %%%%%%%%%%%%%%%

\section*{Acknowledgments}
\paragraph*{Funding:}
This work is supported by the National Key Research and Development Program of China (No. 2021ZD0109902), and the National Natural Science Foundation of China (No. 62275139).
\paragraph*{Author contributions:}
X.L., S.Y., and S.G. conceived the research idea and designed the study. S.Y. carried out the theoretical modeling and numerical simulations. S.Y. and S.G. conducted experiments. C.W., Z.Z., and X.L. contributed to algorithm development and data analysis. H.Z. and X.L. assisted with experimental setup and provided technical discussions. X.L. initiated and supervised the project. All authors discussed the results and contributed to the writing of the manuscript.
\paragraph*{Competing interests:}
There are no competing interests to declare.
\paragraph*{Data and materials availability:}
Data presented in this publication is available on Github with the following link: https://github.com/THPCILab/DMNN. The codes used in the current study are available from the corresponding authors upon reasonable request.

%%%%%%%%%%%%%%%% SUPPLEMENT LIST %%%%%%%%%%%%%%%

% List the contents of your Supplementary Materials, including the numbers of any
% supplementary figures, tables, external data files etc. and any references that are
% cited only in the supplement. In this example, refs. 7-8 are cited only in the supplement.
% Fill out your numbers accordingly and delete any lines that aren't applicable.
\subsection*{Supplementary materials}
Supplementary Notes S1 to S6\\
Figs. S1 to S13\\
Algorithm Tables S1 to S2\\

%%%%%%%%%%%%%%%% END OF MAIN TEXT %%%%%%%%%%%%%%%

\newpage

%%%%%%%%%%%%%%%% START OF SUPPLEMENT %%%%%%%%%%%%%%%

% Figures, tables, equations and pages in the supplement are numbered S1, S2 etc.
\renewcommand{\thefigure}{S\arabic{figure}}
\renewcommand{\thetable}{S\arabic{table}}
\renewcommand{\theequation}{S\arabic{equation}}
\renewcommand{\thepage}{S\arabic{page}}
\setcounter{figure}{0}
\setcounter{table}{0}
\setcounter{equation}{0}
\setcounter{page}{1} % not 0 as \newpage already started a supplementary page
% References continue the numbering from the main text.

%%%%%%%%%%%%%%%% SUPPLEMENT TITLE PAGE %%%%%%%%%%%%%%%

\begin{center}
\section*{Supplementary Materials for\\ \scititle}

% Author list for the supplement
% Indicate the corresponding authors, but do NOT include institutions here
% It would be nice if the template auto-generated this, but doing so is complicated...
	Songtao Yang$^{1,\dagger}$,
	Sheng Gao$^{1,\dagger}$,
	Chu Wu$^{1}$,
	Zejia Zhao$^{1}$,
	Haiou Zhang$^{1}$,
	Xing Lin$^{1,2,\ast}$\and

	\small$^{1}$Department of Electronic Engineering, Tsinghua University, Beijing, 100084, China.\and 
	
	\small$^{2}$Beijing National Research Center for Information Science and Technology, Tsinghua University, Beijing, 100084, China.\and
	
	% Identify at least one corresponding author, with contact email address
	\small$^\ast$Corresponding author. Email: lin-x@tsinghua.edu.cn\and
	
	% Joint contributions can be indicated like this
	\small$^\dagger$These authors contributed equally to this work.
\end{center}

% Fill out the numbers for each type of supplementary material,
% and delete any lines that aren't applicable.
% These are just example numbers that don't match the rest of this template.
\subsubsection*{This PDF file includes:}
Supplementary Notes S1 to S6\\
Figures S1 to S13\\
Algorithm Tables S1 to S2\\

\newpage

%%%%%%%%%%%%%%%% SUPPLEMENTARY TEXT %%%%%%%%%%%%%%%

% ----------- Supplementary Note 1 ------------
\subsubsection*{Supplementary Note 1: Modeling of metasurfaces based on FF-MLP}

Metasurface used in this work integrates the arrays of internal resonated meta-atoms for high-dimensional electromagnetic field modulations. Each meta-atom has three adjustable parameters, $d_a$, $d_b$, and $\theta$, which represent the length of the major axis, the length of the minor axis, and the rotation angle of the ellipse, respectively. The Jones matrix $\mathbf{J}$ is used to describe the high-dimensional modulation. Assuming the input electric field is $\mathbf{E}_{\text{in}} = \begin{bmatrix} E_{ix}, E_{iy} \end{bmatrix}^T$, and the output electric field after modulation is $\mathbf{E}_{\text{out}} = \begin{bmatrix} E_{ox}, E_{oy} \end{bmatrix}^T$, satisfying:
\begin{equation}
\mathbf{E}_{\texttt{out}} = \mathbf{J} \cdot \mathbf{E}_{\texttt{in}}.
\end{equation}

Due to the rotational symmetry geometry of meta-atoms, the Jones matrix of the meta-atom can be written as:
\begin{equation}
\mathbf{J} =
\begin{bmatrix} 
{J}_{xx} &{J}_{xy} \\ 
{J}_{yx} & {J}_{yy} 
\end{bmatrix}=
\begin{bmatrix}
\cos \theta & -\sin \theta \\
\sin \theta & \cos \theta
\end{bmatrix}
\begin{bmatrix}
A_x e^{i\Phi_x} & 0 \\
0 & A_y e^{i\Phi_y}
\end{bmatrix}
\begin{bmatrix}
\cos \theta & \sin \theta \\
-\sin \theta & \cos \theta
\end{bmatrix},
\end{equation}
where $A_x$, $A_y$, $\Phi_x$, and $\Phi_y$ are functions of $d_a$ and $d_b$. Furthermore, we can write the multi-frequency Jones matrix:
\begin{equation}
\mathbf{J}(f) =
\begin{bmatrix} 
\mathbf{J}_{xx}(f) &\mathbf{J}_{xy}(f) \\ 
\mathbf{J}_{yx}(f) & \mathbf{J}_{yy}(f) 
\end{bmatrix}=
\begin{bmatrix}
\cos \theta & -\sin \theta \\
\sin \theta & \cos \theta
\end{bmatrix}
\begin{bmatrix}
\mathbf{A}_x(f) e^{i\mathbf{\Phi}_x(f)} & 0 \\
0 & \mathbf{A}_y(f) e^{i\mathbf{\Phi}_y(f)}
\end{bmatrix}
\begin{bmatrix}
\cos \theta & \sin \theta \\
-\sin \theta & \cos \theta
\end{bmatrix}.
\end{equation}
In accordance with physical laws, the FF-MLP only needs to receive two variables $d_a$ and $d_b$, and outputs $\mathbf{A}_x(f)$, $\mathbf{A}_y(f)$, $\mathbf{\Phi}_x(f)$, and $\mathbf{\Phi}_y(f)$. Combined with the rotation matrix $\mathbf{R}(\theta)$, the multi-frequency Jones matrix at each meta-atom of metasurfaces under different design parameters can be calculated.

The FF-MLP maps the original design parameters to high-dimensional space through a Fourier feature layer, followed by a single fully connected layer to obtain the output results. As illustrated in Supplementary Fig.~\ref{fig:s1}, FF-MLP exhibits a significantly sparser parameter structure compared to conventional deep learning models, enabling more efficient and stable learning. In our network design, the input is a two-dimensional vector $\mathbf{v} = (d_a, d_b)^T$. Fourier feature mapping is a learnable projection matrix $\mathbf{B} \in \mathbb{R}^{M \times 2}$, where $M$ denotes the number of Fourier components (i.e., the mapping size), enabling the model to capture quick variations in the input domain. The input vector is first projected as $\mathbf{z} = 2\pi \mathbf{B} \mathbf{v}$ and the Fourier embedding is then constructed as follows: $\boldsymbol{\gamma}(\mathbf{v}) = 
\begin{bmatrix}
\sin(\mathbf{z}) \\
\cos(\mathbf{z})
\end{bmatrix}
\in \mathbb{R}^{2M \times 1}.$
This high-dimensional feature vector $\boldsymbol{\gamma}(\mathbf{v})$ is subsequently passed through a fully connected hidden layer of dimension $N_{\text{hidden}}$, a ReLU activation function, and finally mapped to the output prediction $S(f) \in \mathbb{R}^{N_f \times 1}$. Therefore, the total number of trainable parameters in a single FF-MLP model is given by: $N_{\text{total}} = 2M + 2MN_{\text{hidden}} + N_{\text{hidden}} + N_{\text{hidden}}N_f + N_f.$
For data acquisition, we traversed $d_a$ and $d_b$ from 1.4~mm to 3.1~mm with a step size of 0.05~mm in CST software, obtaining data from 24~GHz to 32~GHz at intervals of 0.02~GHz under different design parameters, thus the output size of FF-MLP is a $401 \times 1$ vector. Interpolation methods were used to achieve data augmentation, increasing the number of dataset from $35\times35$ to $171\times171$. To balance prediction performance and model compactness, we chose $M = 128$ and $N_{\text{hidden}} = 256$. Therefore, the total number of trainable parameters for a single FF-MLP is: $N_{\text{total}} = 2 \times 128 + 2 \times 128 \times 256 + 256 + 256 \times 401 + 401 = 169105.$

For model validation, we acquired an additional set of responses by sampling $d_a$ in the range of 1.43~mm to 3.03~mm and $d_b$ from 1.48~mm to 3.08~mm, both with a step size of 0.1~mm, resulting in a total of 289 distinct configurations. For comparison, we trained Naive-MLPs comprising two hidden layers each with 256 neurons. The error distributions of the two architectures are shown in Supplementary Fig.~\ref{fig:s2}, while Supplementary Fig.~\ref{fig:s3} and Supplementary Fig.~\ref{fig:s4} present their detailed prediction results. The FF-MLP architecture consistently outperforms the Naive-MLPs in predicting electromagnetic responses.

\newpage

% ----------- Supplementary Note 2 ------------
\subsubsection*{Supplementary Note 2: Forward Inference of DMNN}

Leveraging the high-dimensional nature of microwave optics, the DMNN performs multiple computational tasks in parallel by encoding each task into a distinct dimension of the EM field, such as wavelength or polarization. To enable this functionality, we treat the design parameters $\{\mathbf{D}^{(l)}_{i,j} \in \mathbb{R}^{3} \mid i,j = 1,2,\dots, N;\ l = 1,2,\dots, L\}$ as trainable variables, and use pre-trained FF-MLPs to characterize the high-dimensional EM modulation behavior.

To model the forward inference process, we first simulate the free-space propagation of the EM field components corresponding to each independent dimension using the angular spectrum method (ASM). These components are then coherently combined into a unified high-dimensional field before encountering the modulation layer. For a given input with specific polarization and wavelength, we define the initial electromagnetic field as:
\begin{equation}
\mathbf{E}_{\lambda}^{\text{in}}(x, y) = 
\begin{bmatrix}
E_{\lambda,x}(x, y) \\[2pt]
E_{\lambda,y}(x, y)
\end{bmatrix},
\end{equation}
The modulation at each spatial location is governed by a complex-valued Jones matrix $\mathbf{J}_{\lambda}(x, y) \in \mathbb{C}^{2 \times 2}$, which encodes both amplitude and phase responses under different polarization states. This matrix is inferred from the design parameter vector $\{\mathbf{D}^{(l)}_{i,j}\}$ via pre-trained FF-MLPs:
\begin{equation}
\mathbf{J}^{(l)}_{\lambda}(x, y) = \mathbf{M}_{\lambda}\left( \mathbf{D}^{(l)}_{i,j} \right), \quad 
i = \left\lfloor \frac{x}{a} \right\rfloor + 1,\quad 
j = \left\lfloor \frac{y}{a} \right\rfloor + 1.
\end{equation}
For clarity of notation, we omit the spatial coordinates $(x, y)$ and $(i,j)$ in the following expressions when not ambiguous. With the local modulation matrix defined, the modulated field after the $l$-th layer is computed as:
\begin{equation}
\mathbf{E}_{\lambda,\text{out}}^{(l)} = \mathbf{J}^{(l)}_{\lambda} \cdot \mathbf{E}_{\lambda,\text{in}}^{(l)} = \mathbf{M}_{\lambda}(\mathbf{D}^{(l)}) \cdot \mathbf{E}_{\lambda,\text{in}}^{(l)}.
\end{equation}
\noindent specifically, $\mathbf{E}_{\lambda,\text{in}}^{(1)}=\mathbf{E}_{\lambda}^{\text{in}}$. We then apply the angular spectrum method (ASM) to calculate the field after propagation. The angular spectrum is expressed as:
\begin{equation}
\mathbf{A}_{\lambda}^{(l)}(f_x, f_y) = \mathcal{F}_{x,y}\left\{ \mathbf{E}_{\lambda,\text{out}}^{(l)} \right\},
\end{equation}
and the propagated field at a distance $d$ along the $z$-axis for the next layer is given by:
\begin{equation}
\mathbf H_\lambda(f_x,f_y) =
\begin{cases}
\exp\!\left[j 2\pi d \sqrt{\left(\tfrac{1}{\lambda}\right)^{2} - f_x^{2} - f_y^{2}} \,\right], 
& f_x^{2}+f_y^{2} \le \left(\tfrac{1}{\lambda}\right)^{2}, \\[1.2ex]
0,
& f_x^{2}+f_y^{2} > \left(\tfrac{1}{\lambda}\right)^{2}.
\end{cases}
\end{equation}
\begin{equation}
\begin{split}
\mathbf{E}_{\lambda,\text{in}}^{(l+1)} &= \mathcal{F}^{-1}_{f_x, f_y} \left\{\mathbf{A}_{\lambda}^{(l)}(f_x, f_y) \odot \mathbf H_\lambda(f_x,f_y)\right\}\\
&=\mathcal{F}^{-1}_{f_x, f_y} \left\{\mathcal{F}_{x,y}\left\{ \mathbf{M}_{\lambda}(\mathbf{D}^{(l)}) \cdot \mathbf{E}_{\lambda,\text{in}}^{(l)} \right\} \odot \mathbf H_\lambda(f_x,f_y)\right\}\\
&=\mathbf{U}_{\lambda}\mathbf{M}_{\lambda}(\mathbf{D}^{(l)})\mathbf{E}_{\lambda,\text{in}}^{(l)},
\end{split}
\end{equation}
where $\odot$ denotes the Hadamard (i.e., element-wise) product and $\mathbf{U}_{\lambda}$ is free space propagation operator. By recursively applying the above modulation and propagation steps across all layers, the final high-dimensional EM field at the detection plane can be computed as:
\begin{equation}
\mathbf{E}_{\lambda}^{\text{det}}(\mathbf{D}) = \mathbf{E}_{\lambda,\text{in}}^{(L+1)} = \prod_{l=1}^L {\mathbf{U}_{\lambda}\mathbf{M}_{\lambda}(\mathbf{D}^{(l)})} \mathbf{E}_{\lambda}^{\text{in}}
\end{equation}
\noindent The complete forward inference process is summarized in Algorithm~\ref{alg:forward}.

\newpage

% ----------- Supplementary Note 3 ------------
\subsubsection*{Supplementary Note 3: Improving angle estimation accuracy with output regression}

To account for discretization in practical systems, we initially treat the angle estimation task as a classification problem during training. However, this approach inevitably introduces quantization errors, particularly at the boundaries between adjacent angle classes. Assuming a uniform angular step size of $ \delta $, and a total of 16 discrete detection regions, the full field of view (FOV) becomes $ 16\delta $. For instance, if the angular outputs are uniformly spaced over the interval $ [-7.5\delta, -6.5\delta, \dots, 7.5\delta] $, the theoretical minimum mean absolute error due to quantization is $ 0.25\delta $. We define angular estimation throughput(AET) to assess angular resolution as:
\begin{equation}
AET = \frac{N_d \times N_t \times FOV}{\text{Mean Error}},
\end{equation}
where $N_d$ denotes the number of resolvable angular directions, $N_t$ denotes the number of targets, and $FOV$ is the FOV in degrees. In practice, DMNN often contains rich spatial information beyond the single-peak detection region. Neighboring areas may contain secondary peaks or partially overlapping signals that are often ignored in standard classification schemes. To utilize this additional information, we introduce a refined angle estimation method that incorporates both the maximum and second maximum detected intensity values, $ I_1 $ and $ I_2 $, and their corresponding class numbers $ c_1 $ and $ c_2 $. The estimated angle $ \theta_r $ is then computed by interpolating between the two classes:
\begin{equation}
\theta_r = \frac{c_1 I_1 + \eta c_2 I_2}{I_1 + \eta I_2} \cdot \delta - 8.5\delta + \delta_r,
\end{equation}
where $ \eta $ is a tunable hyperparameter that weights the influence of the second-largest intensity value. When $ \eta = 0 $, the formula reduces to the standard classification output using only the class with largest intensity.

To further enhance angular estimation, we adopt a multi-frequency fusion strategy. In this approach, the angle is estimated independently at multiple incident frequencies $ f_1, f_2, \dots, f_N $, each acting as a weak learner. These individual estimates are then aggregated through weighted averaging to yield a final prediction:
\begin{equation}
\theta = \frac{1}{N} \sum_{i=1}^{N} w_i \theta_i,
\end{equation}
where $ \theta_i $ denotes the estimated angle at frequency $ f_i $, and $ w_i $ is the corresponding weight coefficient. This ensemble-like regression improves accuracy by reducing the variance of single-frequency predictions and leveraging complementary information across spectral channels.

\newpage

% ----------- Supplementary Note 4 ------------
\subsubsection*{Supplementary Note 4: DOA estimation using stage-wise training method}

The training objective of the DMNN is to optimize the spatially distributed design parameters $\{\mathbf{D}^{(l)}_{i,j} \in \mathbb{R}^{3} \mid i,j = 1,2,\dots, N;\ l = 1,2,\dots, L\}$ such that the system performs super-resolution DOA estimation task. During optimization, the design parameters are the only learnable variables. The entire forward inference process, including angular spectrum propagation and field modulation, is differentiable and supports gradient-based optimization through standard automatic differentiation frameworks (e.g., PyTorch).

To ensure stable and task-effective training, we adopt a two-stage optimization strategy. In the first stage, the loss function is defined to align the output field intensity $I_{\lambda, \boldsymbol{p}}(\mathbf{D}) = \left| \boldsymbol{p} \cdot \mathbf{E}^{\mathrm{det}}_{\lambda}(\mathbf{D}) \right|^2$ with the target intensity pattern $\mathrm{TI}_{\lambda, \boldsymbol{p}}$ at the output plane across all dimensions. The objective is written as:
\begin{equation}
\mathcal{L}^{(1)}(\mathbf{D}) = \sum_{\lambda} \sum_{\boldsymbol{p}} \epsilon_{\lambda, \boldsymbol{p}}^{(1)} \cdot \mathrm{MSE}\left( I_{\lambda, \boldsymbol{p}}(\mathbf{D}),\ \zeta_{\lambda, \boldsymbol{p}}^{(1)}\cdot\mathrm{TI}_{\lambda, \boldsymbol{p}} \right),
\end{equation}
\noindent where $\epsilon_{\lambda, \boldsymbol{p}}^{(1)},\ \zeta_{\lambda, \boldsymbol{p}}^{(1)}$ are balancing coefficients.

In the second stage, we emphasize energy within the detection regions. The intensity $I_{\lambda, \boldsymbol{p}}$ is integrated over 16 fixed sub-regions to produce a region-wise energy vector $P_{\lambda, \boldsymbol{p}}(\mathbf{D}) \in \mathbb{R}^{16 \times 1}$, which is compared with the target distribution $\mathrm{TD}_{\lambda, \boldsymbol{p}}$ via:
\begin{equation}
\mathcal{L}^{(2)}(\mathbf{D}) = \sum_{\lambda} \sum_{\boldsymbol{p}} \epsilon_{\lambda, \boldsymbol{p}}^{(2)} \cdot \mathrm{MSE}\left( P_{\lambda, \boldsymbol{p}}(\mathbf{D}),\ \zeta_{\lambda, \boldsymbol{p}}^{(2)}\cdot\mathrm{TD}_{\lambda, \boldsymbol{p}} \right).
\end{equation}
\noindent where $\epsilon_{\lambda, \boldsymbol{p}}^{(2)},\ \zeta_{\lambda, \boldsymbol{p}}^{(2)}$ are balancing coefficients as well.

The full optimization process is summarized in Supplementary Fig.~\ref{fig:s5} and Algorithm~\ref{alg:train}. Through this two-stage learning strategy, the performance of the DMNN in DOA estimation tasks can be effectively enhanced. For low-resolution tasks, which are relatively easier to learn, the performance gain from stage-wise training is limited. However, for high-resolution tasks, conventional one-stage training often struggles to converge to optimal solutions, whereas the stage-wise training method significantly improves the performance (see Supplementary Fig.~\ref{fig:s6} and \ref{fig:s7}). Supplementary Fig.~\ref{fig:s6} shows the convergence of training loss and accuracy of different DOA estimation tasks configured with different angular intervals using stage-wise learning. Supplementary Fig.~\ref{fig:s7} compares the prediction results obtained with and without stage-wise training. For the 1.5$^\circ$ angular interval task, the mean absolute error across the two incident directions is reduced from 0.193$^\circ$ to 0.068$^\circ$. For the 3.0$^\circ$ angular interval task, the error decreases from 0.233$^\circ$ to 0.131$^\circ$. In the 6.0$^\circ$ angular interval case, the error is reduced from 0.505$^\circ$ to 0.362$^\circ$. These results clearly demonstrate the effectiveness of the stage-wise training strategy in improving high-resolution DOA estimation.

\newpage

% ----------- Supplementary Note 5 ------------
\subsubsection*{Supplementary Note 5: DOA estimation of multiple targets by generating incoherent inputs}

Since the DMNN operates within a high-dimensional linear modulation system before electromagnetic field detection, multi-source injection can be analyzed using the superposition principle of optical fields. Considering the case of two incident sources, if the sources occupy different dimensions (e.g., distinct frequencies or polarizations), they can be effectively decoupled and independently processed at the detection stage without mutual interference.

In contrast, when the two sources lie in the same physical dimension, which means they share identical polarization and frequency, there are still two cases. The first case is that they are emitted by the same signal generator, but propagate through different paths before reaching the DMNN, then they typically exhibit high mutual coherence. In such cases, the energy distribution in the output field must be computed based on the interference of their complex fields. The two incident fields must be coherently superposed before calculating the intensity:
\begin{equation}
I_{\text{total}}(x, y) = \left| E_1(x, y) + E_2(x, y) \right|^2.
\end{equation}
However, if the sources are generated by different signal generators, their interference vanishes due to short coherence time. In this case, the total output intensity is simply the sum of the individual intensities:
\begin{equation}
I_{\text{total}}(x, y) = \left| E_1(x, y) \right|^2 + \left| E_2(x, y) \right|^2.
\end{equation}

Supplementary Fig.~\ref{fig:s9} presents the two-target angle estimation results for three different angular intervals in the incoherent case. Supplementary Fig.~\ref{fig:s9}\textbf{a} (top) demonstrates the predicted versus input angles for two x-polarized sources with fixed azimuthal separations of 1.5$^\circ$ and 4.5$^\circ$, and illustrates the estimation error distributions for randomly sampled angle separations. The average estimation errors for the two sources are 0.17$^\circ$ and 0.18$^\circ$. Supplementary Fig.~\ref{fig:s9}\textbf{a} (bottom) demonstrates the corresponding results for elevational angle estimation, the average errors are both 0.17$^\circ$. Supplementary Fig.~\ref{fig:s9}\textbf{b}, \textbf{c} demonstrate the 3.0$^\circ$ and 6.0$^\circ$ angular interval case, with an average estimation error of 0.33$^\circ$ and 0.79$^\circ$, respectively.

\newpage

% ----------- Supplementary Note 6 ------------
\subsubsection*{Supplementary Note 6: Experimental results of all-optical elevational angle estimation}
As a supplement to the experimental results in the main text, this section shows the results of the system's elevational angle estimation using y-polarized electromagnetic field. The experimental system of this work is shown in Supplementary Fig.~\ref{fig:s10}. A representative input with an elevational angle of $2.25^\circ$ was selected with the corresponding input phase distribution shown in Supplementary Fig.\ref{fig:s11}\textbf{a}. The numerically and experimentally measured energy fields at 27, 29, and 31~GHz are shown in Supplementary Fig.\ref{fig:s11}\textbf{b}. All three frequencies exhibit peak energy concentration at the 10th detection region, which is consistent to the design. Supplementary Fig.\ref{fig:s11}\textbf{c} presents the normalized energy distributions over the 16 detection regions for all three frequencies. These correspond to the angular interval midpoints calibrated at 29~GHz. Elevational angle estimation is performed by identifying the maximum and second-maximum values in the energy distributions across the three frequencies. By statistically analyzing the estimated results under different incident angles, the experiment achieved an average estimation error of $0.15^\circ$ within the FOV of [$-11.5^\circ$, $11.5^\circ$], as shown in Fig.~3\textbf{g} of main text. The green curve shows the estimation error for each angle, and the green region indicates the $95\%$ confidence of angle estimation, which is [$-0.41^\circ$, $0.33^\circ$].

\clearpage
\newpage

\begin{algorithm}[th]
\linespread{1.0}\selectfont
\caption{Multi-layer Forward Inference of DMNN}
\label{alg:forward}
\KwIn{Initial optical field $\mathbf{E}_{\lambda,\text{in}}^{(1)}$; \\
\hspace{3.2em} Design parameters $\{\mathbf{D}^{(l)}_{i,j} \in \mathbb{R}^{3} \mid i,j = 1,2,\dots, N;\ l = 1,2,\dots, L\}$; \\
\hspace{3.2em} Propagation distances $d$}
\KwOut{Final high-dimensional optical field $\mathbf{E}_{\lambda}^{\text{det}}$ at the detection plane}
\For{$l \leftarrow 1$ \KwTo $L$}{
    $\mathbf{J}_{\lambda}^{(l)} \leftarrow \mathbf{M}_{\lambda}(\mathbf{D}^{(l)})$ \tcp*[r]{Predict Jones matrix}
    $\mathbf{E}_{\lambda,\text{out}}^{(l)} \leftarrow \mathbf{J}_{\lambda}^{(l)} \cdot \mathbf{E}_{\lambda,\text{in}}^{(l)}$ \tcp*[r]{Apply modulation}
    $\mathbf{A}_{\lambda}^{(l)} \leftarrow \mathcal{F}(\mathbf{E}_{\lambda,\text{out}}^{(l)})$ \tcp*[r]{Angular spectrum (Fourier domain)}
    $\mathbf{A}_{\lambda,\text{in}}^{(l+1)} \leftarrow \mathbf{A}_{\lambda}^{(l)} \odot \mathbf{H}_{\lambda}(f_x,f_y)$ \tcp*[r]{Free-space propagation}
    $\mathbf{E}_{\lambda,\text{in}}^{(l+1)} \leftarrow \mathcal{F}^{-1}(\mathbf{A}_{\lambda,\text{in}}^{(l+1)})$ \tcp*[r]{Back to spatial domain}
}
\KwRet{$\mathbf{E}_{\lambda}^{\mathrm{det}} = \mathbf{E}_{\lambda,\mathrm{in}}^{(L+1)}$}
\end{algorithm}

\clearpage
\newpage

\begin{algorithm}[th]
\linespread{1.0}\selectfont
\caption{Stage-wise Training of DMNN}
\label{alg:train}
\KwIn{Multi-dimensional training dataset $\left\{ \left(\mathbf{E}_{\lambda,\text{in}}^{(1)}, \mathrm{TI}_{\lambda, \boldsymbol{p}}, \mathrm{TD}_{\lambda, \boldsymbol{p}} \right) \right\}$; \\
\hspace{3.2em} Initial design parameters $\mathbf{D}$; \\
\hspace{3.2em} Pre-trained FF-MLPs; \\
\hspace{3.2em} Batch size $B$; \\
\hspace{3.2em} Task weights $\{\epsilon_{\lambda, \boldsymbol{p}}^{(1)}, \zeta_{\lambda, \boldsymbol{p}}^{(1)}\}$, $\{\epsilon_{\lambda, \boldsymbol{p}}^{(2)}, \zeta_{\lambda, \boldsymbol{p}}^{(2)}\}$ for Stage 1 and 2; \\
\hspace{3.2em} Epochs $T_1$, $T_2$, learning rate $lr_1$, $lr_2$ for Stage 1 and 2}
\KwOut{Optimized design parameters $\mathbf{D}^*$}
Randomly initialize $\mathbf{D}$ ;
\BlankLine
\textbf{Stage 1: Intensity Alignment}\;
\For{$t \leftarrow 1$ \KwTo $T_1$}{
    \ForEach{minibatch of $B$ samples}{
        \ForEach{$\lambda, \boldsymbol{p}$}{
            \For{$b \leftarrow 1$ \KwTo $B$}{
                Propagate $\mathbf{E}^{(1)}_{\lambda,\text{in}, b}$ through DMNN to obtain $\mathbf{E}_{\lambda,b}^{\text{det}}(\mathbf{D})$ \;
                $I_{\lambda, \boldsymbol{p},b}(\mathbf{D}) \leftarrow \left| \boldsymbol{p} \cdot \mathbf{E}^{\mathrm{det}}_{\lambda,b}(\mathbf{D}) \right|^2$ \;
            }
            $\mathcal{L}^{(1)}_{\lambda,\boldsymbol{p}} \leftarrow \frac{1}{B} \sum_{b=1}^B \mathrm{MSE}\left( I_{\lambda, \boldsymbol{p},b}(\mathbf{D}),\ \zeta_{\lambda, \boldsymbol{p}}^{(1)}\cdot\mathrm{TI}_{\lambda, \boldsymbol{p},b} \right)$ \;
        }
        $\mathcal{L}^{(1)} \leftarrow \sum_{\lambda} \sum_{\boldsymbol{p}} \epsilon_{\lambda, \boldsymbol{p}}^{(1)}\cdot \mathcal{L}_{\lambda,\boldsymbol{p}}^{(1)}$ \;
        Update $\mathbf{D}$ via Adam optimizer \;
    }
}
\BlankLine
\textbf{Stage 2: Region-wise Refinement}\;
\For{$t \leftarrow 1$ \KwTo $T_2$}{
    \ForEach{minibatch of $B$ samples}{
        \ForEach{$\lambda,\boldsymbol{ p}$}{
            \For{$b \leftarrow 1$ \KwTo $B$}{
                Propagate $\mathbf{E}^{(1)}_{\lambda,\text{in}, b}$ through DMNN to obtain $\mathbf{E}_{\lambda,b}^{\text{det}}(\mathbf{D})$ \;
                $I_{\lambda, \boldsymbol{p},b}(\mathbf{D}) \leftarrow \left| \boldsymbol{p} \cdot \mathbf{E}^{\mathrm{det}}_{\lambda,b}(\mathbf{D}) \right|^2$ \;
                $\mathbf{P}_{\lambda, \boldsymbol{p}, b} \leftarrow$ integrate $I_{\lambda, \boldsymbol{p},b}$ over 16 detection regions \;
            }
            $\mathcal{L}^{(2)}_{\lambda,\boldsymbol{p}} \leftarrow \frac{1}{B} \sum_{b=1}^B \mathrm{MSE}\left( \mathbf{P}_{\lambda, \boldsymbol{p}, b},\ \zeta_{\lambda, \boldsymbol{p}}^{(2)}\cdot\mathrm{TD}_{\lambda, \boldsymbol{p}, b} \right)$ \;
        }
        $\mathcal{L}^{(2)} \leftarrow \sum_{\lambda} \sum_{\boldsymbol{p}} \epsilon_{\lambda, \boldsymbol{p}}^{(2)} \cdot \mathcal{L}_{\lambda,\boldsymbol{p}}^{(2)}$ \;
        Update $\mathbf{D}$ via Adam optimizer \;
    }
}

\KwRet{$\mathbf{D}^* \leftarrow \mathbf{D}$}
\end{algorithm}
\clearpage
% If your supplement is very short you might need to uncomment the following line to avoid
% layout problems with the figures and tables.
%\newpage

%%%%%%%%%%%%%%%% SUPPLEMENTARY FIGURES %%%%%%%%%%%%%%%
\newpage

\begin{figure}
\centering
\includegraphics[width=\linewidth]{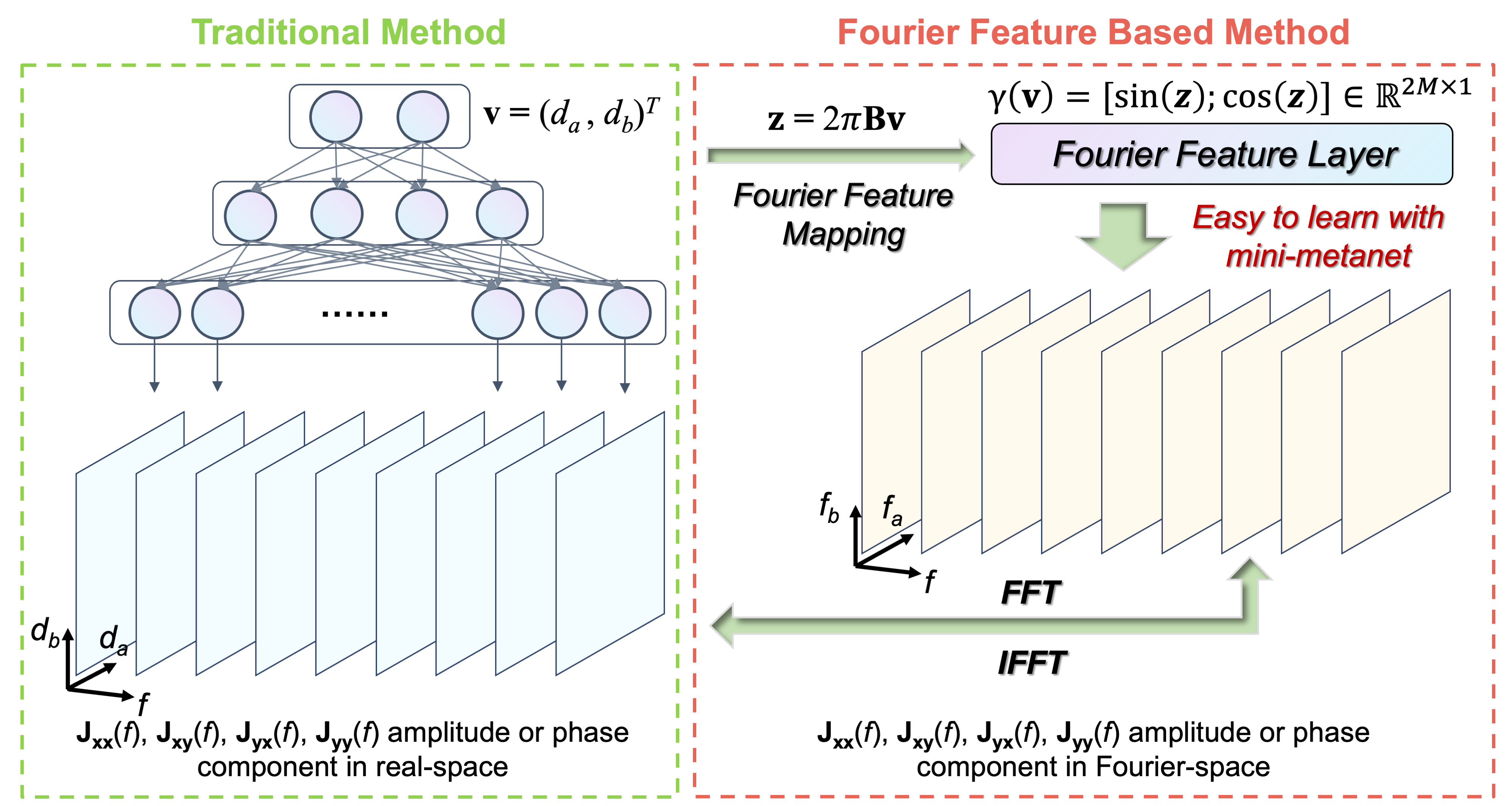}
\caption{\textbf{Comparison of traditional deep learning method and Fourier feature based method.}}
\label{fig:s1}
\end{figure}

\begin{figure}
\centering
\includegraphics[width=\linewidth]{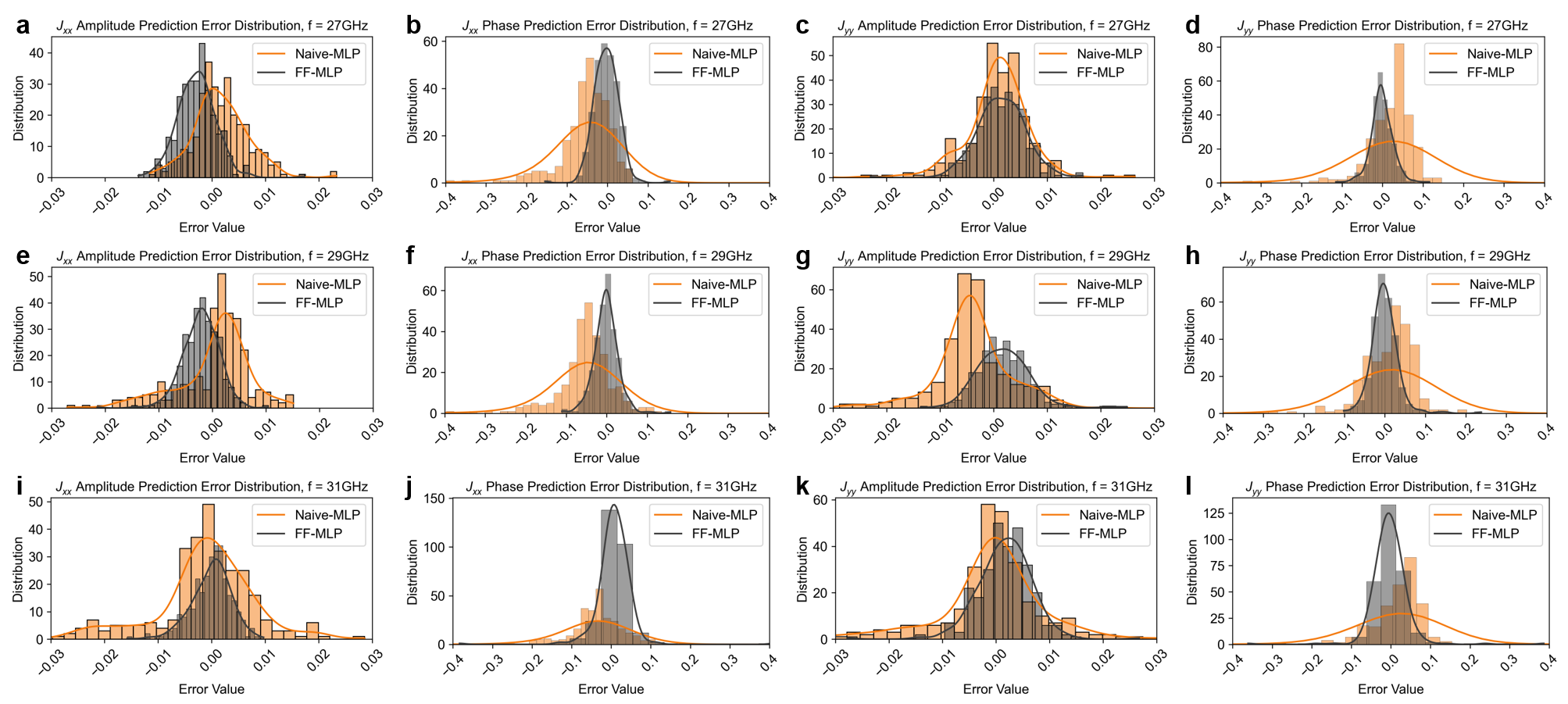}
  \caption{\textbf{Error distribution of different models at different frequency.} \textbf{a-d} $J_{xx}$ amplitude, $J_{xx}$ phase, $J_{yy}$ amplitude and $J_{yy}$ phase prediction at 27~GHz, \textbf{(e-h)} 29~GHz and \textbf{(i-l)} 31~GHz.}
\label{fig:s2}
\end{figure}

\begin{figure}
\centering
\includegraphics[width=\linewidth]{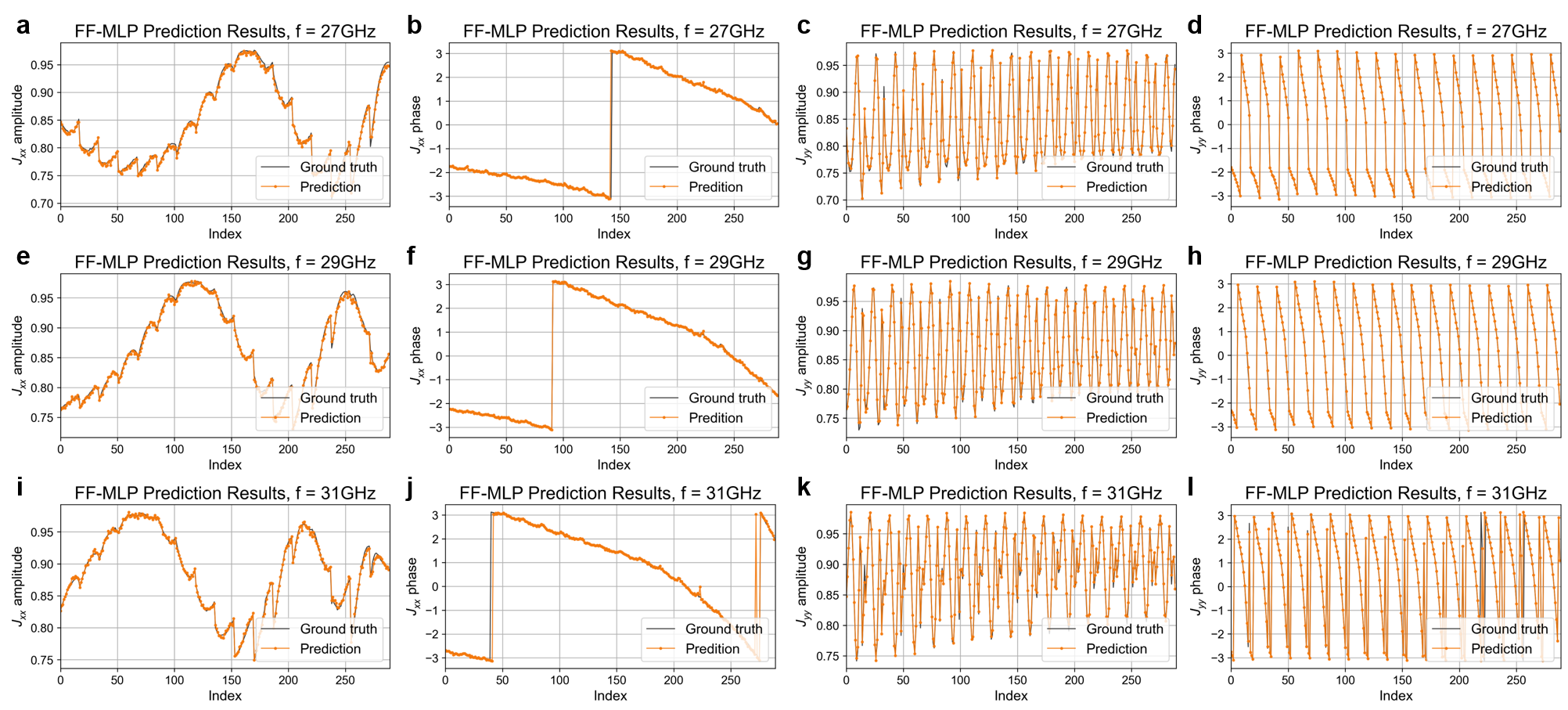}
\caption{\textbf{Prediction result of FF-MLP at different frequency.}  \textbf{a-d} $J_{xx}$ amplitude, $J_{xx}$ phase, $J_{yy}$ amplitude and $J_{yy}$  phase prediction at different design parameters at 27~GHz, \textbf{(e-h)} 29~GHz and \textbf{(i-l)} 31~GHz.}
\label{fig:s3}
\end{figure}

\begin{figure}
\centering
\includegraphics[width=\linewidth]{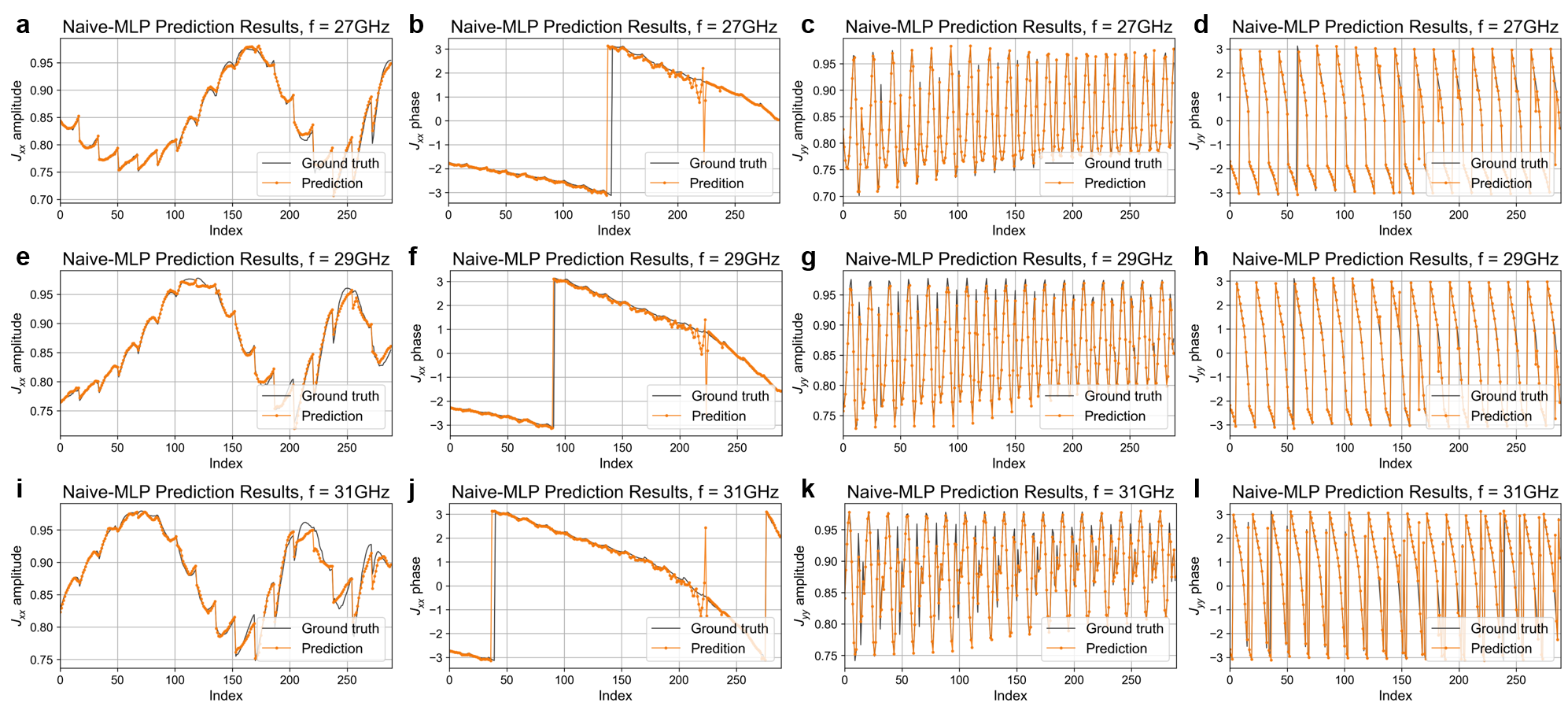}
\caption{\textbf{Prediction result of Naive-MLP at different frequency.} \textbf{a-d} $J_{xx}$ amplitude, $J_{xx}$ phase, $J_{yy}$ amplitude and $J_{yy}$ phase prediction at different design parameters at 27~GHz, \textbf{(e-h)} 29~GHz and \textbf{(i-l)} 31~GHz.}
\label{fig:s4}
\end{figure}

\begin{figure}
\centering
\includegraphics[width=0.9\linewidth]{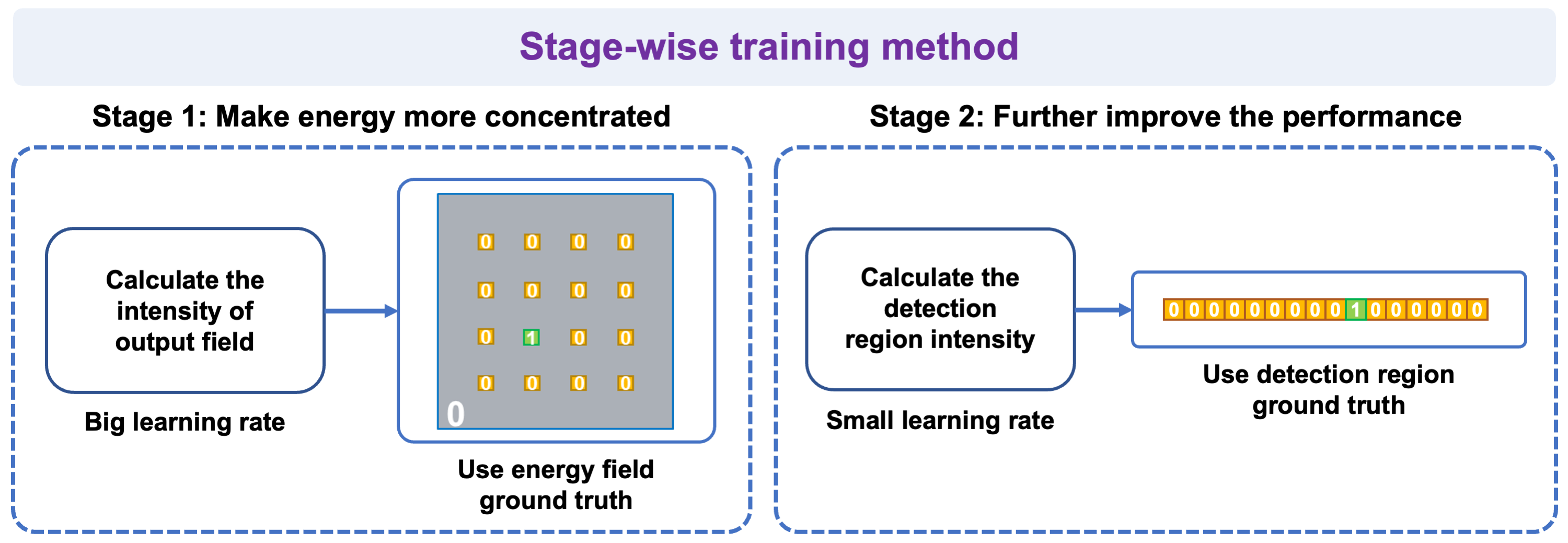}
\caption{\textbf{Stage-wise training method for improving the performance of DMNNs.}}
\label{fig:s5}
\end{figure}

\begin{figure}
\centering
\includegraphics[width=0.7\linewidth]{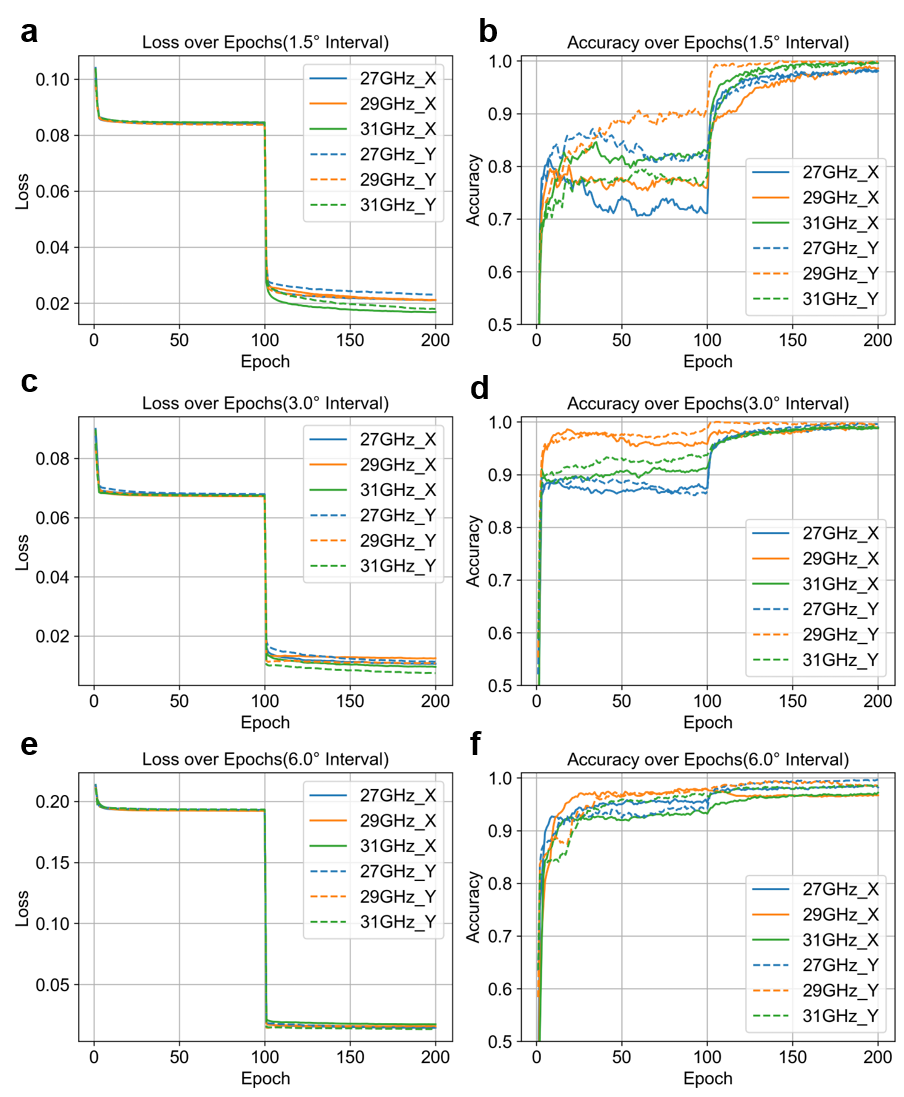}
\caption{\textbf{The convergence of training loss and accuracy of different angle estimation tasks configured with different angular intervals using stage-wise learning.} \textbf{a,c,e} present the training loss curves, and \textbf{(b,d,f)} show the corresponding accuracy curves for classification tasks with angular intervals of $1.5^\circ$, $3.0^\circ$, and $6.0^\circ$, respectively. Each line represents a distinct frequency, i.e., 27~GHz, 29~GHz, or 31~GHz, and a certain x- or y-polarization state (X, Y).}
\label{fig:s6}
\end{figure}

\begin{figure}
\centering
\includegraphics[width=\linewidth]{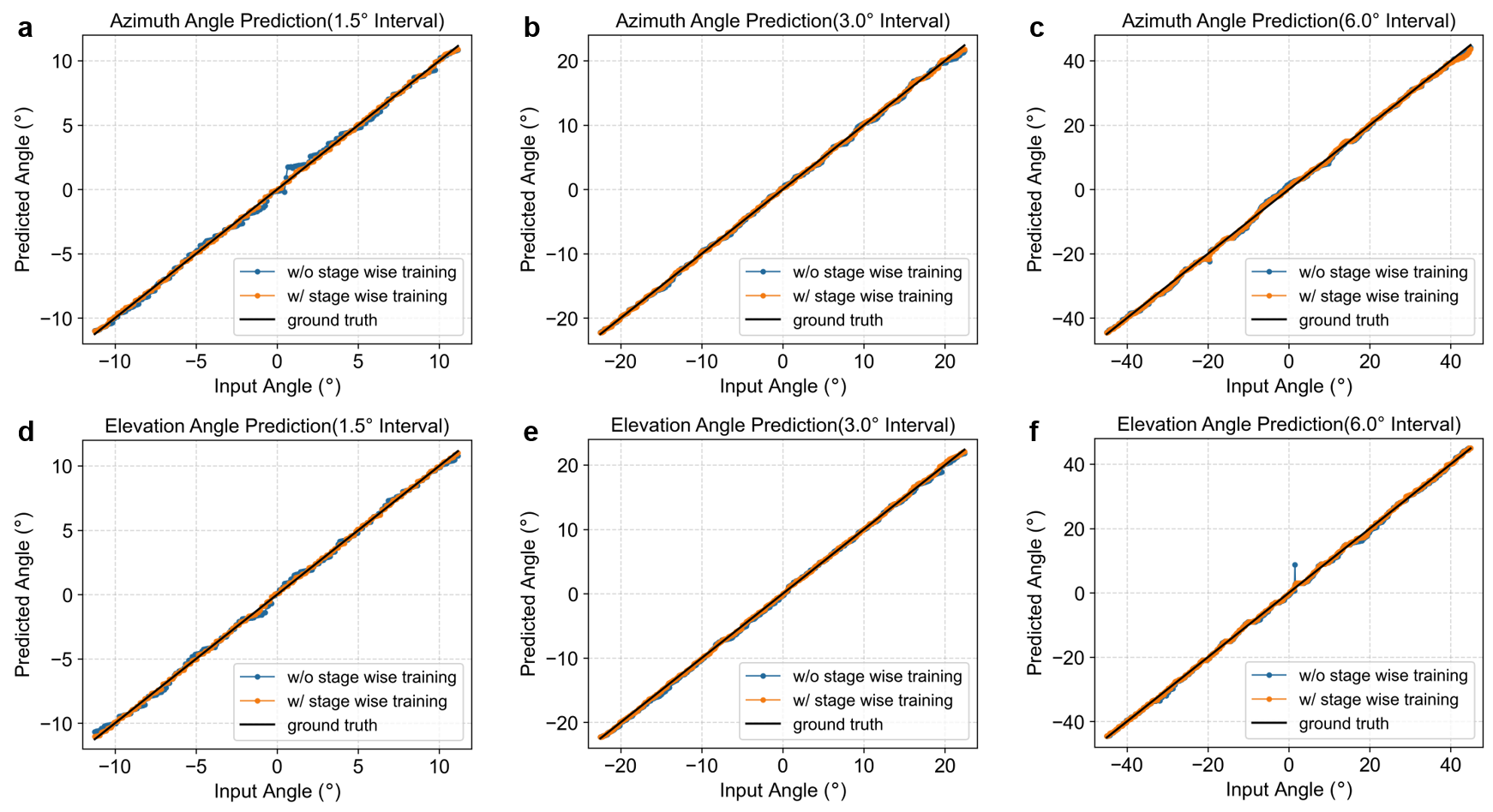}
\caption{\textbf{Numerical results of azimuthal and elevational angle estimation under varying angular intervals, with and without stage-wise training.} \textbf{a-c} display the azimuthal angle prediction results, and \textbf{d-f} show the elevational angle prediction results, across tasks with $1.5^\circ$, $3.0^\circ$, and $6.0^\circ$ angular intervals, respectively. The orange curves represent predictions from models trained using stage-wise learning, while the blue curves correspond to models trained without stage-wise training. The black diagonal line denotes ground truth target angles.}
\label{fig:s7}
\end{figure}

\begin{figure}
\centering
\includegraphics[width=\linewidth]{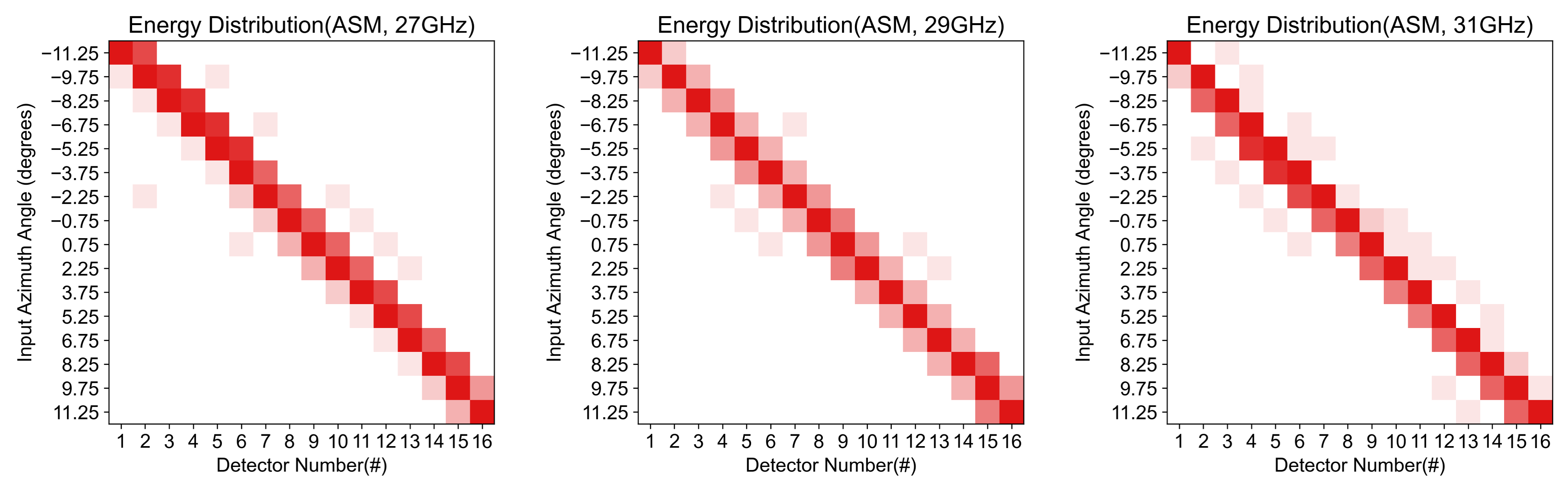}
\caption{\textbf{Numerical energy distribution results of all-optical DMNN for super-resolution azimuthal angle estimation of a single target.}}
\label{fig:s8}
\end{figure}

\begin{figure}
\centering
\includegraphics[width=\linewidth]{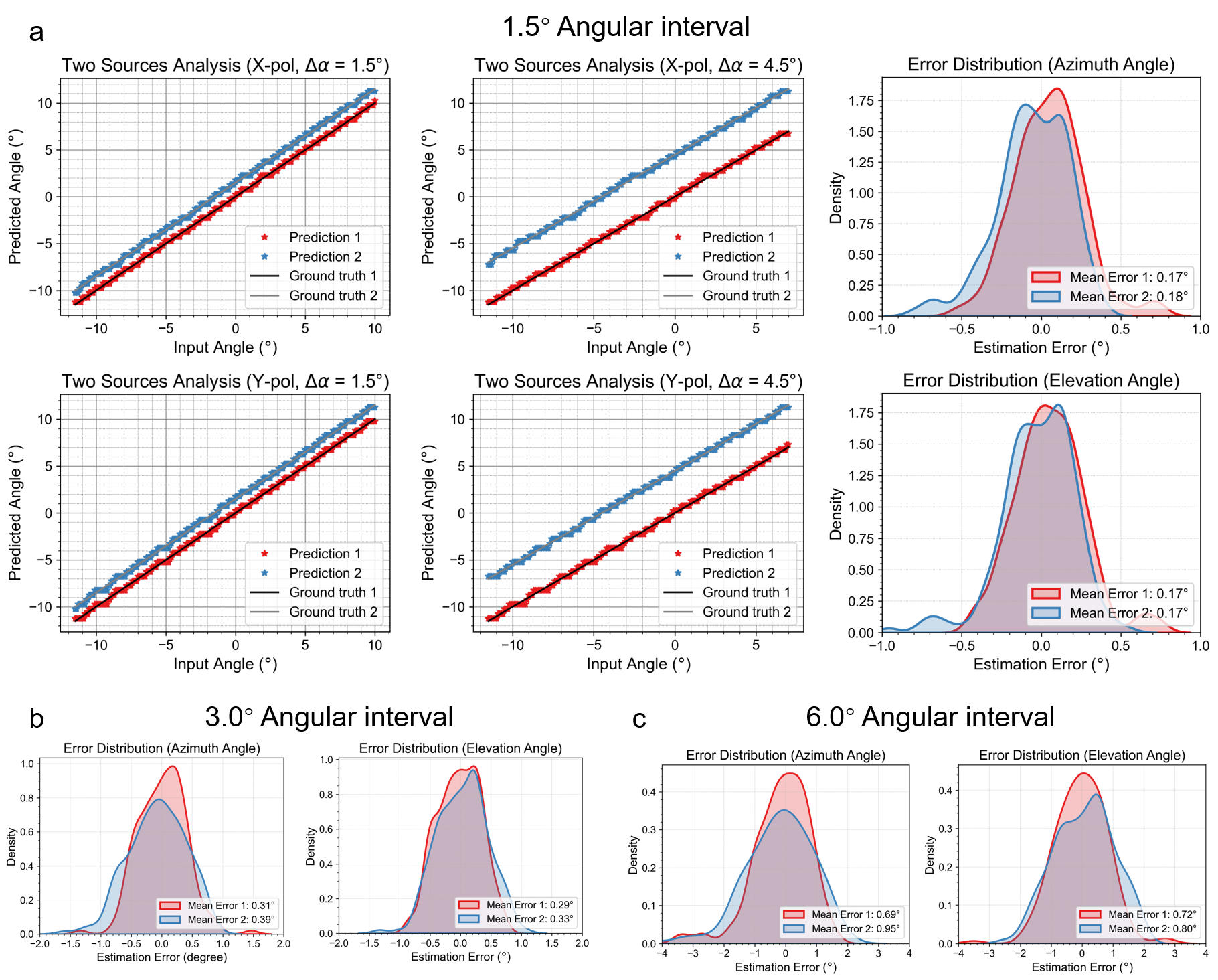}
\caption{\textbf{Numerical results of all-optical DMNN for two-target angle estimation and error distributions at various angular intervals.}}
\label{fig:s9}
\end{figure}

\begin{figure}
\centering
\includegraphics[width=\linewidth]{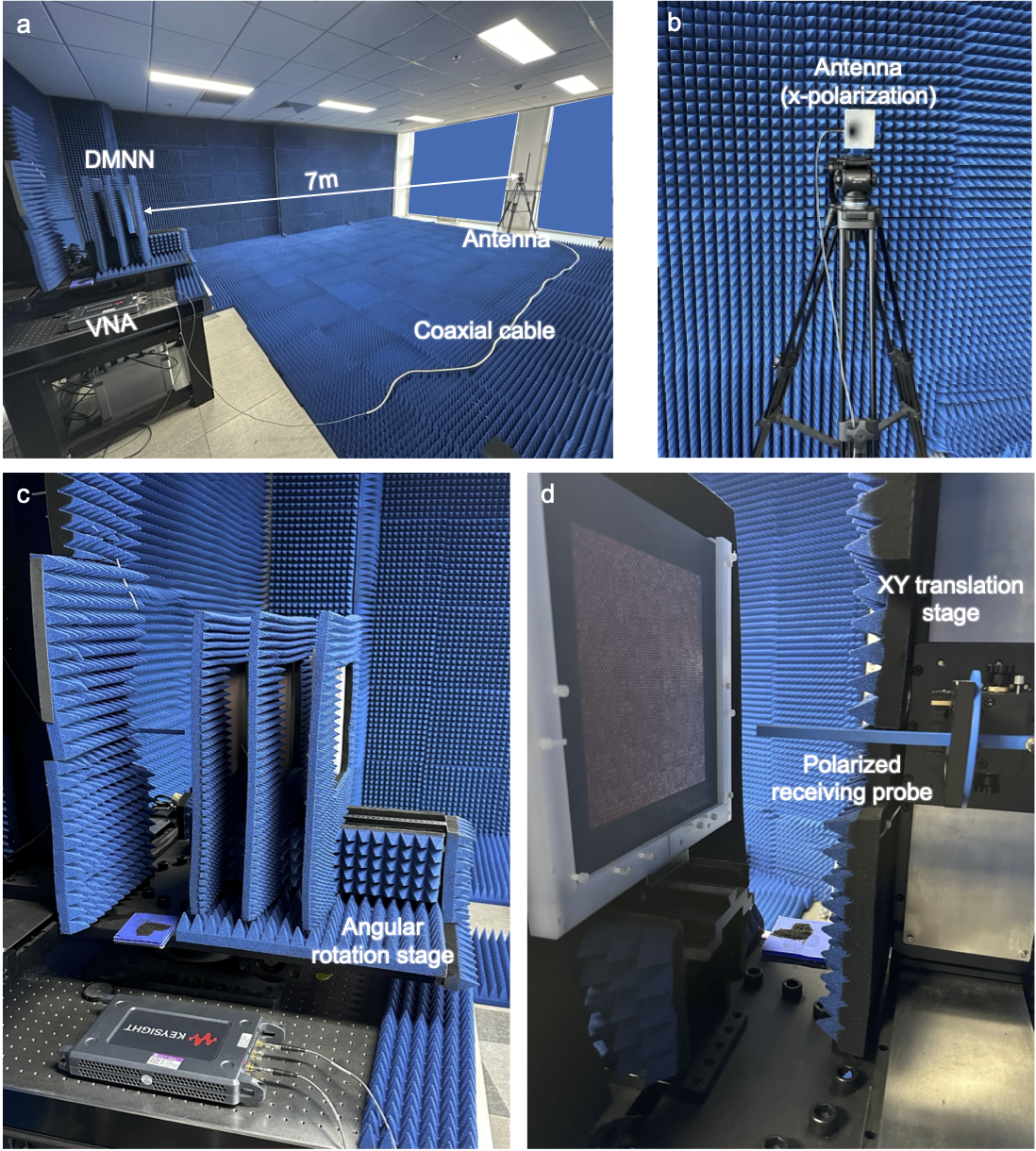}
\caption{\textbf{Experimental setup for testing DMNNs.} \textbf{a} Overview of the microwave anechoic chamber with the VNA, coaxial cable, antenna, and DMNN under test. \textbf{b} Transmitting horn antenna aligned for x-polarization. \textbf{c} Angular rotation stage used to adjust the incident angle. \textbf{d} Polarized receiving probe mounted on the XY translation stage.}
\label{fig:s10}
\end{figure}

\begin{figure}
\centering
\includegraphics[width=\linewidth]{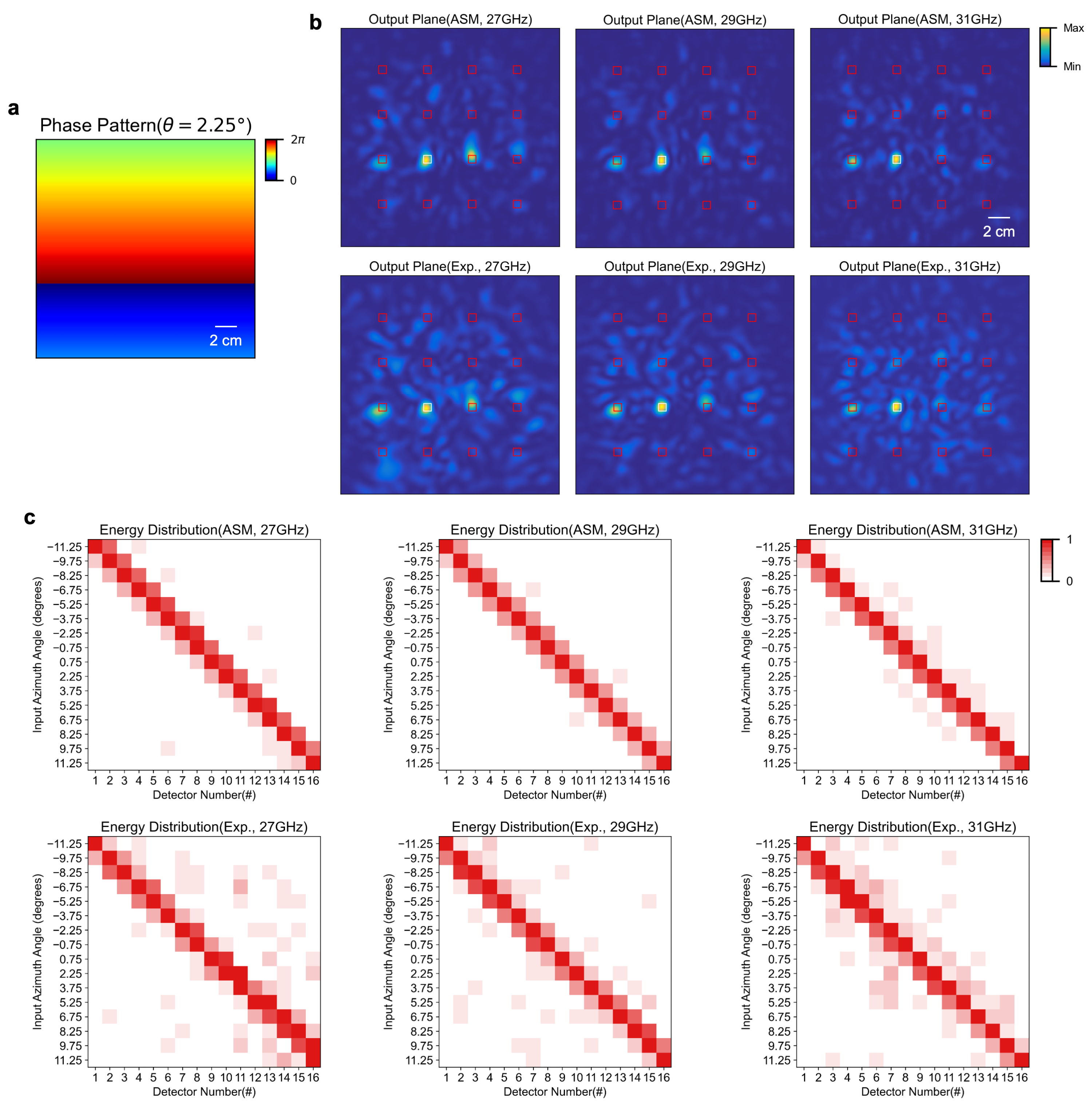}
\caption{\textbf{Numerical and experimental results of all-optical DMNN for super-resolution elevational angle estimation of a single target.} \textbf{a} Y polarization incidence for elevational angle estimation (e.g. $2.25^\circ$ incidence). \textbf{b} Simulated and experimental energy field of 27, 29 and 31~GHz. \textbf{c} Simulated and experimental energy distribution across detectors under 16 incident elevational angles at 27~GHz, 29~GHz, and 31~GHz, with the angles chosen as the midpoints for the 29~GHz case.}
\label{fig:s11}
\end{figure}

\begin{figure}
\centering
\includegraphics[width=\linewidth]{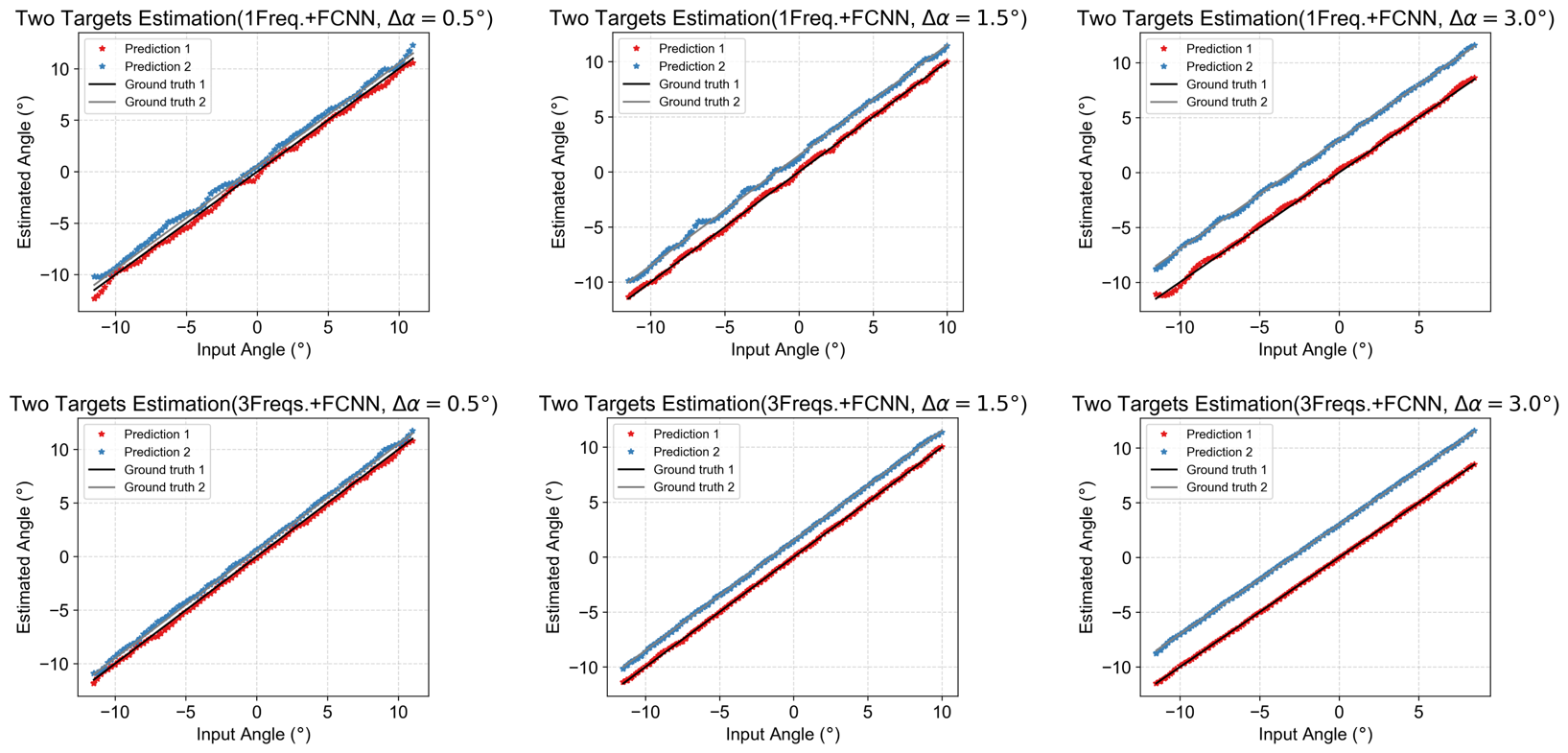}
\caption{\textbf{Experimental results of optoelectronic DMNN for two-target azimuthal angle estimation at $0.5^\circ$, $1.5^\circ$ and $3.0^\circ$ separation.}}
\label{fig:s12}
\end{figure}

\begin{figure}
\centering
\includegraphics[width=\linewidth]{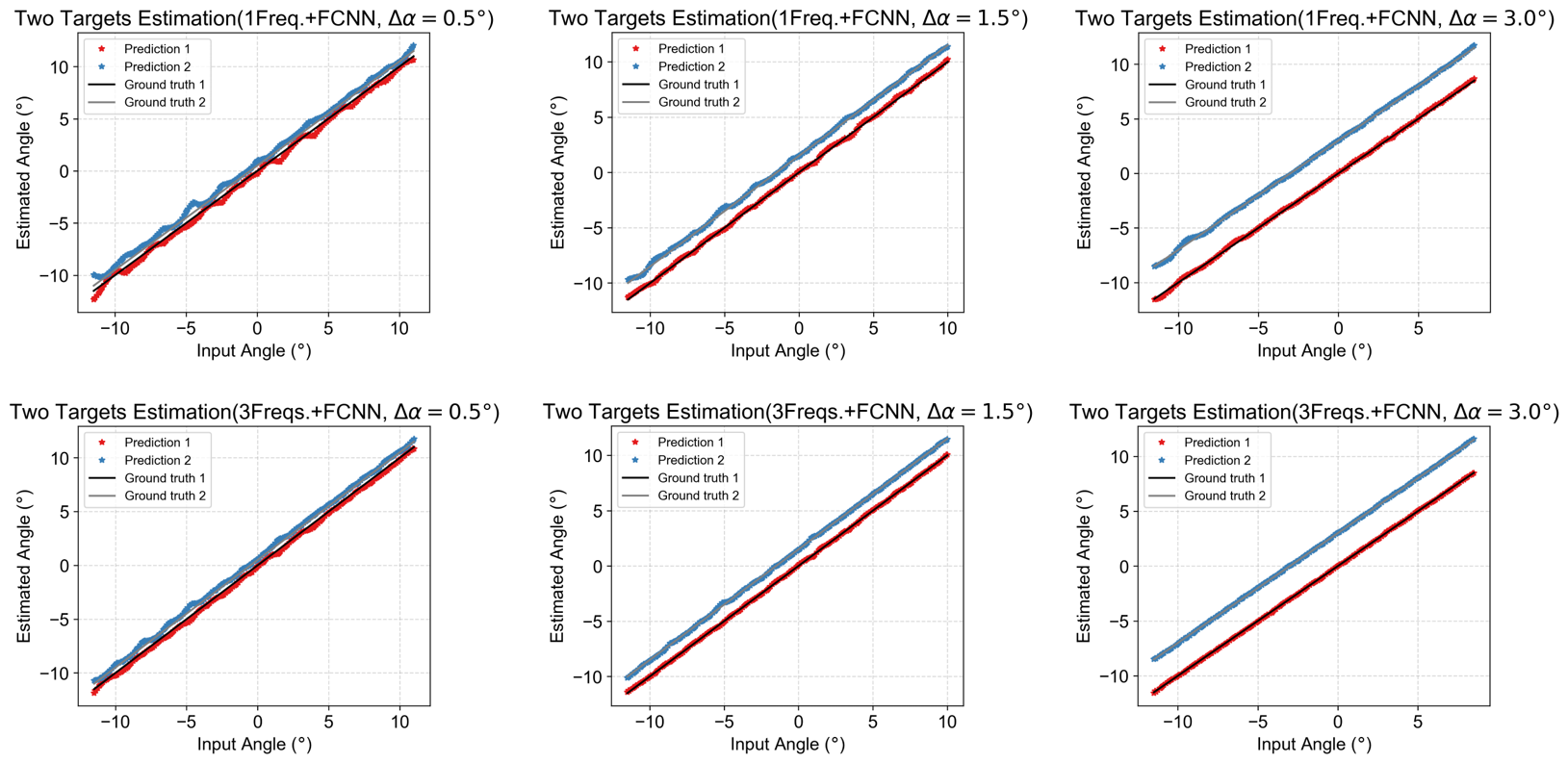}
\caption{\textbf{Experimental results of optoelectronic DMNN for two-target elevational angle estimation at $0.5^\circ$, $1.5^\circ$ and $3.0^\circ$ separation.}}
\label{fig:s13}
\end{figure}

\end{document}